\documentclass[aps,a4paper,reprint,showpacs,amsmath,amssymb,superscriptaddress,floatfix]{revtex4-1}

\usepackage{amssymb}
\usepackage{latexsym}
\usepackage[dvips]{graphicx}
\usepackage{epsfig}
\usepackage{dcolumn}
\usepackage{amsmath}
\usepackage{amsfonts}
\usepackage{bm}
\usepackage{color}

\usepackage{hyperref}
\hypersetup{colorlinks,linkcolor=blue,urlcolor=blue,citecolor=blue}

\newcommand{\be}{\begin{equation}}
\newcommand{\ee}{\end{equation}}
\newcommand{\bea}{\begin{eqnarray}}
\newcommand{\eea}{\end{eqnarray}}

\newcommand{\p}{\partial}
\newcommand{\s}{\sigma}

\renewcommand{\vec}[1]{{\bm #1}}

\begin{document}
\title{Physics of Arbitrary Doped Kondo Lattices: from \\
        a Commensurate Insulator to a Heavy Luttinger Liquid and a Protected Helical Metal}

\author{A. M. Tsvelik}
\affiliation{Condensed Matter Physics and Materials Science Division, Brookhaven National Laboratory, Upton, NY 11973-5000, USA}

\author{O. M. Yevtushenko}
\affiliation{Ludwig Maximilian University, Arnold Sommerfeld Center and Center for Nano-Science,
             Munich, DE-80333, Germany}

\date{\today}

\begin{abstract}
We study one-dimensional Kondo Lattices (KL) which consist of itinerant electrons
interacting with Kondo impurities (KI) - localized quantum magnetic moments. We focus
on KL with isotropic exchange interaction between electrons and KI and with a high
KI density. The latter determines the principal r{\'o}le of the indirect interaction between
KI for the low energy physics. Namely, the Kondo physics becomes suppressed
and all properties are governed by spin ordering. We present a first-ever comprehensive
analytical theory of such KL at an arbitrary doping and predict a variety of regimes with
different electronic phases.
They range from commensurate insulators (at filling factors 1/2, 1/4 and 3/4) to metals
with strongly interacting conduction electrons (close to these three special cases) to an
exotic phase of a helical metal. The helical metals can
provide a unique platform for realization of an emergent protection of ballistic transport
in quantum wires. We compare out theory with previously obtained
numerical results and discuss possible experiments where the theory could be tested.
%%%
% \\ \\
% Subject Area: Condensed Matter Physics (Strongly Correlated Materials, Magnetism)
%%%
\end{abstract}


\maketitle

\section{Introduction}

Kondo lattice is a dense D-dimensional array of local quantum moments (Kondo impurities) interacting
with conduction electrons. KL have been intensively studied during the past two decades in
different regimes and contexts, starting from physics of the Kondo effect and magnetic systems
to topological insulators and the emergent protection of the ideal transport, see
Reviews \cite{tsunetsugu_1997,review-gulacsi,shibata_1999} and Refs.
\cite{doniach_1977,read_1984,auerbach_1986,fazekas_1991,sigrist_1992,tsunetsugu_1992,troyer_1993,ueda_1993,Tsvelik_1994,Shibata_1995,ZachEmKiv,shibata_1996,shibata_1997,Honner_1997,sikkema_1997,mcculloch_2002,xavier_2002,white_2002,novais_2002,Novais_2002b,Juozapavicius_2002,xavier_2003,xavier_2004,yang_2008,smerat_2011,peters_2012,MaciejkoLattice,aynajian_2012,AAY_2013,Yevt-Helical,khait_2018}.
%%%
% \cite{doniach_1977,read_1984,auerbach_1986,fazekas_1991,sigrist_1992,tsunetsugu_1992,troyer_1993,ueda_1993,Tsvelik_1994,Shibata_1995,ZachEmKiv,shibata_1996,shibata_1997,Honner_1997,sikkema_1997,mcculloch_2002,xavier_2002,white_2002,novais_2002,Novais_2002b,xavier_2003,xavier_2004,yang_2008,peters_2012,MaciejkoLattice,aynajian_2012,AAY_2013,Yevt-Helical,TsvYev_2015,Schimmel_2016,khait_2018}.
%%%
There is a long standing question whether the physics of KL
%%%
% (dense arrays of local moments interacting with conduction electrons)
%%%
may resemble that of solitary magnetic impurities in a nonmagnetic host.
The model does follow this scenario for $ D \rightarrow \infty $ \cite{georges_1996}.
At intermediate values of $D>1$ the physics of KL is believed to be determined by the competition between the Kondo screening
and the Ruderman-Kittel-Kosuya-Yosida (RKKY) \cite{Kittel} interaction, as illustrated by the famous Doniach's
phase diagram \cite{doniach_1977}. It has been suggested that,
if the RKKY interaction wins, the system orders magnetically or, perhaps, becomes some a kind of spin liquid.
In one dimension (1D), where long range magnetic order does not occur, things may be more interesting. The physics
of the 1D KLs is the subject of the present paper.

The Doniach's criterion states that the RKKY interaction wins in 1D when the distance
between the spins is smaller then a crossover distance:
\[
  \xi_s < \xi_{\rm cr} \sim \xi \Big[ \vartheta(E_F) J_K^2/T_K\Big]^{1/2} \, ,
\]
where $ \xi $ is the lattice constant, $J_K$ is the exchange integral, $\vartheta(E_F)$ is the
density of states at the Fermi level and $T_K$ is the Kondo temperature. If $ \vartheta(E_F)J_K
<1 $ and the Coulomb interaction is absent (or weak)  then $ \vartheta(E_F) J_K^2/T_K \gg 1 $
and $ \xi_{\rm cr} \gg \xi $. The range $ \xi \lesssim \xi_s \ll \xi_{\rm cr} $ corresponds to the dense KL
whose physics is dominated by RKKY. It is the regime we are interested in this paper.

\begin{figure}

   \includegraphics[width=0.45 \textwidth]{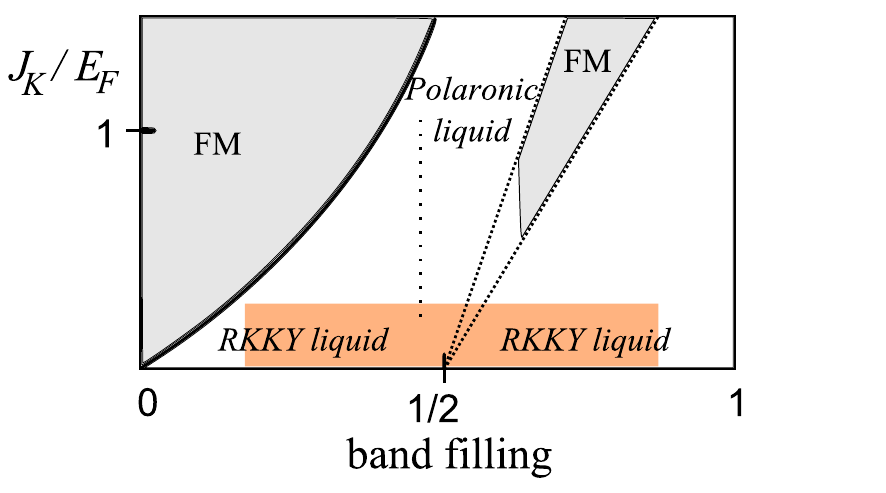}

\vspace{0.3cm}

   \includegraphics[width=0.4 \textwidth]{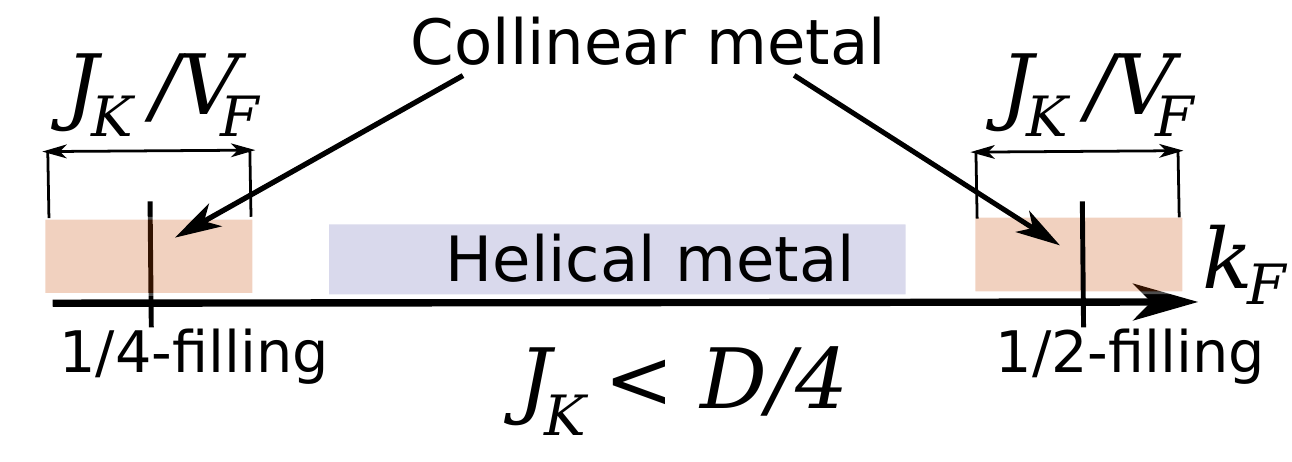}
% \vspace{-0.25cm}
   \includegraphics[width=0.4 \textwidth]{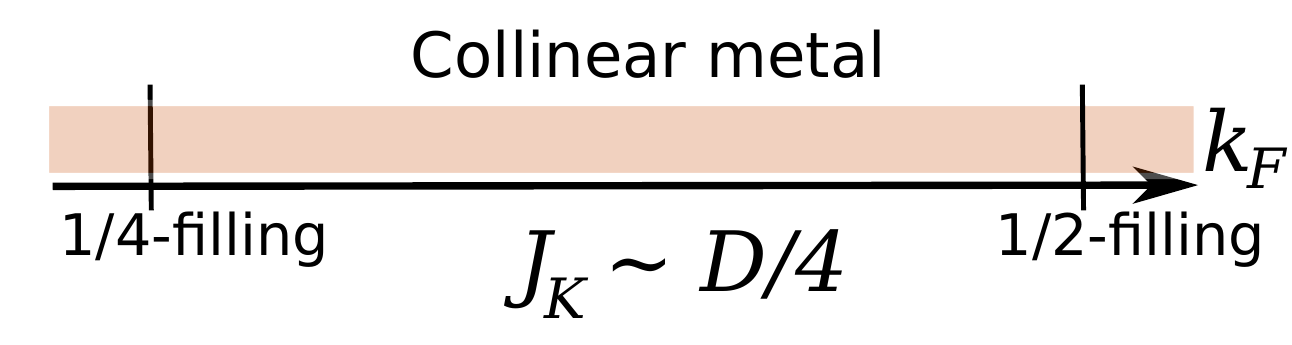}

\vspace{-0.25cm}
   \caption{
        \label{GS-NumPhDiagr} (color on-line)
        The upper panel: a cartoon illustrating the weak and the moderate coupling regimes of the phase diagram
         obtained numerically for 1D KL in Ref.\cite{mcculloch_2002}. The two shaded regions are ferromagnetic
        phases. The region called ``Polaronic Liquid'' corresponds to the large Fermi surface whose volume incorporates both conduction and localized electrons. The Kondo
        effect is suppressed in all regions. The orange box marks the range of parameters considered
        in the present paper.
        The central and the lower panels: phase diagrams of 1D KL  obtained  for small $ J_K/E_F $, see
        explanations in Sect.\ref{SecSummary}
           }
\end{figure}

One of the first results for a rotationally invariant 1D KL was obtained by one of us
as early as 1994 \cite{Tsvelik_1994}. It was shown that, in the 1D KL with a high density of KI at half
filling and relatively small Kondo coupling, $ J_K  \ll D, E_F $ ($ D $ is the band width), there is
really no competition: the RKKY interaction always overwhelms the Kondo screening  and the physics
is governed by the electron backscattering from the short range antiferromagnetic fluctuations.
%%%
% Later, it has been realized by one of us \cite{Tsvelik_1994} that, at least for 1D KL at half-filling with a high density
% of KI and relatively small Kondo coupling, $ J_K $, there is really no competition: RKKY overwhelms the Kondo effect
% and the physics is governed by the spin ordering.
%%%
Numerical results of Ref.\cite{mcculloch_2002} confirm the absence of the Kondo effect for much larger range
of parameters.
%%%
% For this case the numerical calculations indicate that the RKKY interaction always dominates.
%%%
% The ferromagnetic phase fully dominates at large ratio of the Kondo exchange to the electron Fermi energy,
% $ J_K / E_F > 1$.
%%%
For strong coupling,  $ J_K / E_F > 1$,  the ferromagnetism dominates. At smaller values of $ J_K / E_F \le 1 $,
there are two paramagnetic regions separated by a narrow ferromagnetic one, see the upper panel of Fig.\ref{GS-NumPhDiagr}.
The position of the maximum in the momentum-dependent structure factor of the spins is different above and
below this intermediate ferromagnetic region.  At larger $ J_K $, the maximum is located at $ \pi/\xi - 2k_F$, with
$ k_F $ being the Fermi momentum of the band electrons. Such peak position corresponds to the scenario
where both the local moments and the conduction electrons contribute to the Fermi surface (FS)  volume
(the so-called large FS). On the other hand,
at small $ J_K $, the maximum is at $ 2 k_F $.  It was suggested in Ref.\cite{mcculloch_2002} that this corresponds to
the small FS. However, we will argue that the state at a generic filling is $ 4 k_F $-Charge Density Wave (CDW) with
short range spin fluctuations centered at $2k_F$. The Friedel oscillations at $4k_F$ do not distinguish between large
and small FS. The large FS has been also found in the recent Density Matrix Renormalization
Group study conducted  away from 1/4, 3/4  and 1/2 fillings for the relatively large $ J_K $, Ref.\cite{khait_2018}. The
authors of this paper have reported existence of a heavy Tomonaga-Luttinger liquid (TLL) with gapless charge and
spin excitations and Friedel oscillations at $k_F^* = \pi/2\xi - k_F$.
%%%
% The position of the maximum in the momentum-dependent spin structure factor indicates that this ferromagnetic region
% separates ground states with large and small FSs. At larger $ J_K $, the maximum is located at $ \pi - 2k_F$, with
% $ k_F $ being the Fermi momentum, which corresponds to the scenario where both KIs and delocalized electrons
% contribute into FS (large FS). At small $ J_K $, the maximum is at $2k_F$ (small FS). The large FS state has been
% recently studied numerically in Ref.\cite{khait_2018}.
%%%

We will argue
%%%
% in the present paper
%%%
that the RKKY interaction generically dominates in 1D dense KLs. For KLs with magnetic
anisotropy, it has been  demonstrated in our previous publications \cite{Schimmel_2016,TsvYev_2015} for the case of
incommensurate filling. Ref.\cite{Schimmel_2016} contains Renormalization Group arguments which demonstrate
suppression of the Kondo effect.

%%%
% argument that, in 1D,  the RKKY interaction governs the physics.
%%%
The Kondo effects can dominate in the 1D KL
only under some specific conditions, e.g. for small concentrations of local moments, strongly broken SU(2)
spin-rotation symmetry and strong Coulomb
interactions \cite{Yevt_2018} or strongly enlarged symmetry (SU($N$) instead of SU(2) with $ N \gg 1 $) \cite{shibata_1997}.

The surest indication of the RKKY-dominated physics is that it is insensitive to the
sign of $ J_K $ but, at the same time, is very sensitive to doping and anisotropy of the
Kondo coupling. For example:
%%%
% , as we will demonstrate,
%%%
% , if the coupling is isotropic, then
%%%
(i) KL at half-filling is an insulator with gapped or critical spin excitations;  (ii)  at quarter-filling,
KL is also an insulator but with a strong tendency for spin dimerization,
which agrees with the numerical study reported in Ref.\cite{xavier_2003};
%%%
% Spin dimerization has been observed in numerical study of KL at quarter-filling \cite{xavier_2003};
%%%
(iii)~at incommensurate filling, KL with generalized SU(N) symmetry for large $ N \gg 1 $ has been predicted
to be TLL \cite{shibata_1997};
%%%
% (iv) recent Density Matrix Renormalization Group study has detected a heavy TLL in KL at a generic
% (neither half- nor quarter-) filling \cite{khait_2018}.
%%%
% At the same time,
%%%
(iv)~the anisotropic incommensurate KL is also described by TLL either with a
spin-charge separation (easy-axis anisotropy) or with a separation of different helical sectors
(easy-plane anisotropy) \cite{TsvYev_2015,Schimmel_2016}. The latter phase
is characterized by broken helical symmetry of fermions which governs a partial protection of ballistic
transport against effects of disorder and localization.  We remind the readers that helicity
of 1D electrons is defined as the sign of the product of their spin and chirality.

We will consider the isotropic case with SU(2) symmetry and, thus, complete the picture
of the RKKY dominated physics in the dense 1D KL.
Many of the prominent features of 1D KL have been  observed only in numerical studies.
%%%
% since the existing analytical approaches could address only some special cases.
%%%
%  look like  disconnected pieces of a puzzle and
%%%
% Certainly, this calls for creating a generalizing and comprehensive approach.
%%%
{\it The goal of the present paper}
is to develop an analytical approach for the region of $ J_K / E_F \ll 1 $,  see Fig.\ref{GS-NumPhDiagr},
where numerical calculations are difficult to perform, but analytical methods become very powerful.
%%%
%  We have concentrated on 1D, dense, and SU(2)-symmetric KLs and,
%%%
% {\it The goal of the present paper} is to
% develop such an approach. We have chosen to study 1D, dense, and SU(2)-symmetric KLs. We concentrate
% exclusively on the case with small values of $J/\epsilon_F$ where numerical approaches become difficult but
% analytical methods are very powerful, see Fig.\ref{GS-NumPhDiagr}.
%%%
By utilizing various field-theoretical methods, we have developed a  fully analytical and controlled description
of the dense 1D KLs at arbitrary doping. We will show below that 1D KL can form three distinct phases: (i) the insulator at special
commensurate fillings, or (ii) a usual metal formed by interacting electrons when the band filling is close to the special
commensurate fillings, or (iii) $ 4 k_F $-charge-density-wave (CDW) state with gapped spin excitations. Remarkably, the
third phase can also be described as a  metal where the  transport is carried by helical Dirac fermions.
We have determined conditions under which one or another conducting phase appears. In particular, interacting spinful
fermions (either a Fermi liquid or TLL) always exist close to the commensurate insulator and form a usual 1D metal. If the
Kondo coupling is relatively large, this phase exists at any generic filling and becomes the heavy TLL.

The $ 4 k_F $-CDW phase with helical transport can exist only at $ J_K \ll E_F$ far from the commensurate insulator.
Therefore, it can be detected only when parameters are properly tuned. On the other hand, it is very notable
because it possesses an emergent protection against disorder and localization. The emergent protection caused
by interactions is well known and attracts ever-growing attention of theoreticians
\cite{pershin_2004,braunecker_2009a,braunecker_2009b,braunecker_2010,kloeffel_2011,klinovaja_2011a,klinovaja_2011b,klinovaja_2012,kainaris_2015,TsvYev_2015,Schimmel_2016,pedder_2016,kainaris_2017}, and experimentalists \cite{quay_2010,scheller_2014,kammhuber_2017,heedt_2017}. To the best of our knowledge,
all previously known examples were found in the systems with broken (either spontaneously or at the level of the
Hamiltonian) SU(2)-symmetry. Our finding is novel from this point of view because we predict such a protection
to appear in the rotationally invariant system.
%%%
% with broken local discrete $ Z_2 $ helical symmetry.
%%%
% This suggests that the SU(2) symmetry is not the main ingredient
%
% it possesses an emergent protection against disorder and localization without breaking the SU(2)-symmetry.
%%%

We note in passing that, a competition of the RKKY interaction with a direct Heisenberg exchange in dense 1D KL
introduces additional level of complexity and can  lead
to appearance of exotic phases with a nontrivial spin order, such as a chiral spin liquid \cite{KH-CSL} or a chiral lattice
supersolid \cite{TI-SupSol}.

The rest of the paper is organized as follows: Section \ref{SecModel} describes  the model used in our
study. Separation of fast and slow variables is explained in Section \ref{VarSep}.
Section \ref{SecSummary} contains a brief and non-technical summary of our results at the level
of mean-field approximation. Section \ref{QMTheor} is more technical as it is devoted to the detailed field
theoretical description of all phases in KL.
%%%
% Readers, who are not interested in details of the fully quantum mechanical theory, can jump over these Sections and read
%%%
In Section \ref{SecExper}, we discuss possible numerical studies and experiments related to our theory.
Section \ref{SecConcl} contains conclusions. Technical details are presented in Appendices.

The mean field analysis and the protected transport in the $ 4 k_F $-CDW phase are addressed in detail
in Ref.\cite{KL-SU2-short}. In the present paper, we concentrate mainly on the quantum theory which
allows us to analyze the phase diagram of the KL. Repetition of results reported in Ref.\cite{KL-SU2-short}
is reduced here to a minimum necessary to make  the paper self-contained.

\section{The model \label{SecModel}}

We start from the standard KL Hamiltonian:
\bea
  \label{model}
  \hat{H} = \sum_n  \Big[
             & - & t \left( \psi^\dagger_{n}\psi_{n+1} + h.c. \right) - \mu \, \psi^\dagger_{n} \psi_{n} + \\
             & + & J_K\psi_n^\dagger (\vec\s, \vec{S}_n) \psi_n
                                \Big],
  \nonumber
\eea
where $ \psi_n \equiv \{ \psi_{n,\uparrow}, \psi_{n,\downarrow} \}^{\rm T} $ \ are electron annihilation
($ \psi^\dagger_n $ - creation) spinor operators;  ${\vec S}_n$ are quantum  spins with magnitude
$ s $; $ \vec\s \equiv \{ \s_x, \s_y, \s_z \} $ are Pauli matrices in the spin space;
%%%
% $ J_K $ is the Kondo coupling constant;
%%%
$ t $ and $ \mu $ are the electron hopping and the chemical potential, respectively;
summation runs over lattice sites. For simplicity, we do not distinguish constants of KI and
crystalline lattices, $ \xi_s = \xi $. We also assume that $ s J_K \ll D $ where $ D = 2 t $
and consider only low temperatures, $ T \to 0 $.
%%%
% is the band width.
%%%

\section{Separating slow \\
              and fast variables \label{VarSep}}

We are going to derive an effective action for the low energy sector of the theory.
%%%
% We are going to derive an effective low energy theory described in terms of the effective action.
%%%
The crucial step is to single out smooth modes. It is technically convenient to restrict
ourselves to the case $ | \mu |,~|J_K| \ll t $ which allows us to linearize the dispersion
relation and introduce right- and left moving fermions, $ \psi_\pm $, in a standard way
\cite{Giamarchi}. In the continuous limit, the fermionic Lagrangian density takes the form
\be
   \label{Lf}
   {\cal L}_F[\psi_\pm] = \sum_{\nu=\pm} \psi^\dagger_\nu \p_\nu \psi_\nu \, ; \quad
   \p_\pm \equiv \p_\tau \mp i v_F \p_x \, .
\ee
Here $ v_F $ is the Fermi velocity, $ \nu $ is the chiral index which indicates the direction
of motion, $ \p_\nu $ is the chiral derivative and $ \tau $ is the imaginary time.

Following Refs.\cite{Tsvelik_1994,TsvYev_2015,Schimmel_2016}, we keep in the electron-KI
interaction only backscattering terms which are the most relevant in the case of the dense
1D KL. The part of the electron-KI interaction  describing the  backscattering on the site $ n $
reads as:
%%%
% The Lagrangian, which describes backscattering on the site $ n $, reads as:
%%%
\bea
\label{Lbs}
   {\cal L}_{\rm bs}(n) & = & J_K \left[ R^\dagger_n (\vec\s, \vec{S}_n ) L_n e^{-2 i k_F x_n} + h.c \right] ; \\
   R & \equiv & \psi_+,  \, L \equiv \psi_- \, ; \ x_n \equiv n \xi .
\nonumber
\eea
%%%
% with $ k_F $ being the Fermi momentum. The continuous limit of $ \hat{H}_{\rm bs} $ will be taken after the next step.
%%%
% We emphasize that, since there is no Coulomb interaction, throwing
% away the forward scattering does not mean neglecting the Kondo physics.
% If there were only one KI backscattered fermions could be unfolded producing
% the Kondo forward scattering with the spin flip.
%%%
At large inter-impurity distance, the backscattering with a spin flip is a part of  the Kondo screening physics.
However, as we show below, the physics of dense KL is quite different. This will be proven by
insensitivity of all answers to $ {\rm sign} (J_K) $. The Kondo screening  is suppressed in our
model if $ T_K \ll J_K \ll v_F / \xi $. The second inequality is important to accomplish
separation of the fast and the slow modes \cite{TsvYev_2015,Schimmel_2016}.

$ {\cal L}_{\rm bs}(n) $ contains fast $ 2 k_F $-oscillations which must be absorbed into the
spin configuration.  We perform this step in the path integral approach where the spin operators
are replaced by integration over a normalized vector field. We decompose this field as
%%%
% This step can be performed by using a normalized vector field for the
% spin degrees of freedom and applying decomposition
%%%
\bea
   \label{SpinDecomp}
   \vec{S}_n/s = \vec{m} + b \Big(
             & \vec{e}_1 & \, \cos(\alpha) \cos(q x_n + \theta) + \\
             & \vec{e}_2 & \, \sin(\alpha) \sin(q x_n + \theta)
                                          \Bigr)  \sqrt{1 -  \vec{m}^2 } \, .
   \nonumber
\eea
%%%
% Note that $ \vec{m} $ is related to the Kondo physics while $ \vec{e}_{1,2} $
% govern the gap physics.
%%%
where $ q \simeq 2 k_F $, $ \theta $ is a constant (coordinate independent) phase shift;
$ \{ \vec{e}_1, \vec{e}_2, \vec{m} \} $ with $ | \vec{e}_{1,2} | = 1 $ is an orthogonal triad
of vector fields  whose coordinate dependence is smooth on the scale $ 1 / k_F $. The constant $ b $ and
the angle $ \alpha $ must be chosen to solve normalization $ | \vec{S}/s | = 1 $. Eq.(\ref{SpinDecomp})
is generic and it allows only for three possible choices of $ b $, $ \alpha $ and $ \theta $ determined
by the band filling: either 1/2, $ q x_n = 2 k_F x_n = \pi n \Rightarrow b = 1, \, \theta = \alpha = 0 $,
or 1/4, $ q x_n = 2 k_F x_n = \pi n / 2 \Rightarrow b = \sqrt{2}, \ \theta = \pi/4 $, or a generic one,
$ b = \sqrt{2}, \ \theta = 0, \ \alpha = \pi/4 \ $ \cite{KL-SU2-short}.

Using the machinery of advanced field-theoretical methods becomes easier if the vectors $ \vec{e}_{1,2} $
are expressed via a matrix $ g \in \mbox{SU(2)} $
\be
\label{BasisFromSU2}
   \vec{e}_{1,2,3} = \frac{1}{2} {\rm tr}[ \vec{\sigma} g \sigma_{x,y,z} g^{-1}]  \, ;
\ee
see \appendixname~\ref{UslRel}. $ g $ is a smooth function of $ x $ and $ \tau $ and it
governs new rotated fermionic basis
\bea
 \label{RotatedFerm}
 \tilde{R} & \equiv & g^{-1} R , \ \tilde{L} \equiv g^{-1} L \, ; \\
  {\cal L}_F[\tilde{R},\tilde{L}] & = & \tilde{R}^\dagger ( \p_+ + g^{-1} \p_+ g ) \tilde{R} +
                                                           \tilde{L}^\dagger ( \p_- + g^{-1} \p_- g ) \tilde{L}.
 \nonumber
\eea
Jacobian of this rotation is given in  \appendixname~\ref{SU2-Jacobian}.

Now we insert Eq.(\ref{SpinDecomp}) into Eq.(\ref{Lbs}), select non-oscillatory  parts
of $ {\cal L}_{\rm bs} $ for three above mentioned cases, and take the continuous limit.
This yields the smooth part of the Lagrangian density:
\bea
    \label{half_params}
     {\cal L}_{\rm bs}^{(1/2)} \! & = & \! \tilde{J} \sqrt{1 -  \vec{m}^2 }
                                          \left( \tilde{R}^\dagger \s_x \tilde{L} + h.c \right) ; \\
    \label{quart_params}
     {\cal L}_{\rm bs}^{(1/4)} \! & = & \! \frac{\tilde{J} \sqrt{1 -  \vec{m}^2 }}{\sqrt{2}} \Bigl( e^{i \pi / 4}
                      \tilde{R}^\dagger [ \cos(\alpha) \s_x + i  \sin(\alpha) \s_y] \tilde{L} + \cr
                      & & \qquad \qquad \qquad + \, h.c \Bigr) ; \\
  \label{hel_params}
     {\cal L}_{\rm bs}^{\rm (gen)} \! & = & \! \tilde{J} \sqrt{1 -  \vec{m}^2 }
                                          \left( \tilde{R}^\dagger \s_{-} \tilde{L} + h.c \right) .
\eea
Here the superscript of $ \cal{L}_{\rm bs} $ denotes the band filling; $ \tilde{J} \equiv s J_K / 2 $, and
$ \s_\pm = (\s_1 \pm i \s_2) / 2 $. Note that the low energy physics of KLs with the 1/4- and 3/4-filling is
equivalent in our model. Therefore, we often discuss only quarter-filling in the text and do not repeat the
same discussion for the case of the 3/4-filling.

One can give the following interpretations to the above introduced spin configurations:
Eq.(\ref{half_params}) corresponds to a staggered configuration of spins at half-filling, $ \uparrow \downarrow $,
which was studied in Ref.\cite{Tsvelik_1994}. Eq.(\ref{quart_params}) assumes  two spins up- two spins down
configuration, $ \uparrow \uparrow \downarrow \downarrow $, which agrees with the spin dimerization tendency
observed numerically in Ref.\cite{xavier_2003} at quarter-filling. Eq.(\ref{hel_params}) is a rotationally invariant
counterpart of the helical spin configuration discovered in Refs.\cite{TsvYev_2015,Schimmel_2016} in the anisotropic
KL at incommensurate fillings. A simplified version of the spin configuration Eq.(\ref{hel_params}) was used also in
Ref.\cite{fazekas_1991} for analyzing the phase diagram of KL at the level of the mean field approximation.
We emphasize, however, that our approach is more advanced and generic since it has allowed us to go
much further, namely, to derive the low energy effective action and to take into account quantum fluctuations for
all phases.

\begin{figure}[t]
\vspace{-0.5cm}
   \includegraphics[width=0.225 \textwidth]{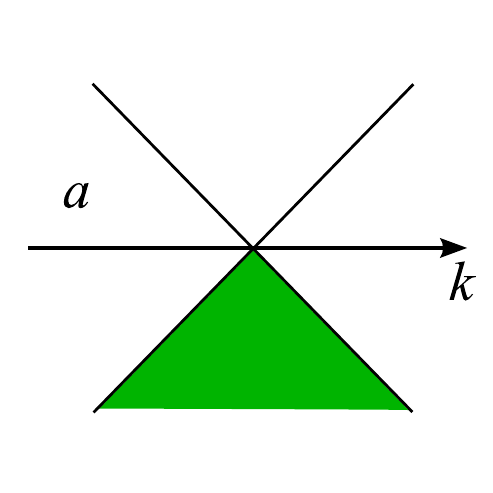}
   \includegraphics[width=0.225 \textwidth]{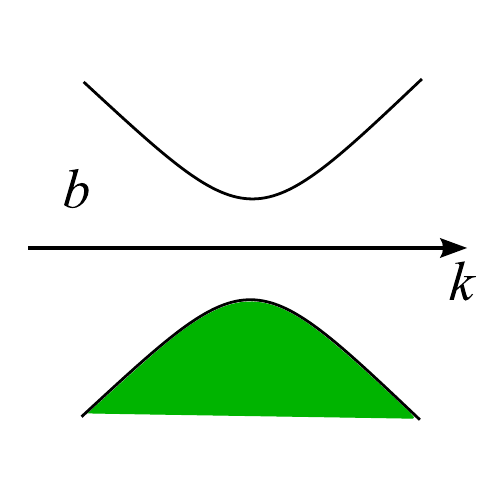}

\vspace{-1.0cm}
   \includegraphics[width=0.225 \textwidth]{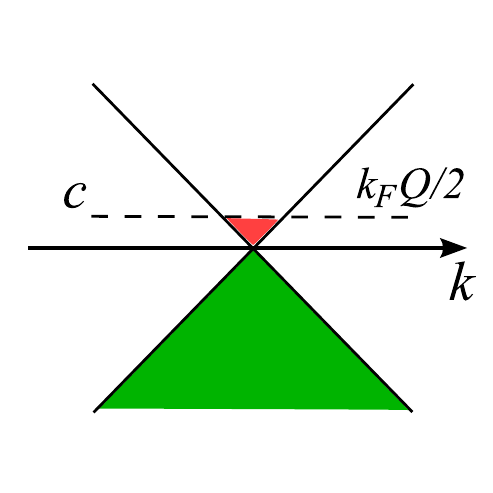}
   \includegraphics[width=0.225 \textwidth]{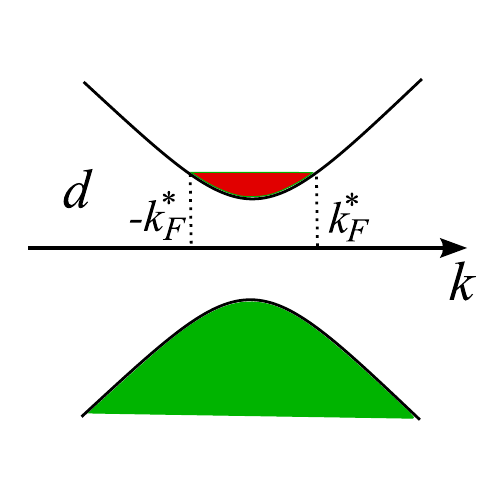}

\vspace{-1.0cm}
   \includegraphics[width=0.225 \textwidth]{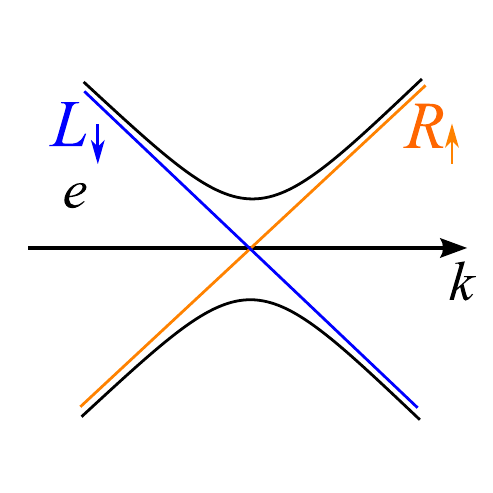}

\vspace{-0.5cm}
   \caption{
        \label{GS-Fig1} (color on-line)
        Panels (a) and (b): the ground state of fermions $ \tilde{R}, \tilde{L} $ with
        zero chemical potential (green region) before and after opening of the gap
        at the level of $ \mu $. Panels (c) and (d): the ground state of the fermions
        for the case of a finite positive chemical potential.
        The red regions mark the states between the gap and the chemical potential which
        can form Tomonaga-Luttinger liquid. Panel (e)
        exemplifies a co-existence of the gapped (black lines) and the gapless
        (orange and blue lines) helical fermions.
           }
\end{figure}

Backscattering of the Dirac fermions $ \{ \tilde{R}, \tilde{L} \} $ opens a gap, $ \Delta \propto \tilde{J}
\sqrt{1 -  \vec{m}^2 }$, in their spectrum. If the gap is opened at (or close to) the Dirac point, defined
by the level of the chemical potential, it decreases the ground state energy of the fermions see
Fig.\ref{GS-Fig1}a--d. The larger the gap, the stronger is the gain in the fermionic energy:
\be
   \label{DeltaEnGS}
   \delta E_{\rm GS} = - \vartheta_0 \, \xi \sum_{k=1,2} \Delta^2_{k} \log\bigl( D / | \Delta_{k} | ) ;
\ee
see Eqs.(\ref{Gap12}--\ref{GapHel}) below and Ref.\cite{KL-SU2-short}. Here, $ \vartheta_0 = 1 / \pi v_F $
is the density of states of the 1D Dirac fermions and the sum runs over two fermionic (helical) sectors.
Since the spin degrees of freedom do not have kinetic
energy the minimum of the ground state energy is reached at the maximum of the fermionic gap.
This indicates that $ | \vec{m} | $ is the gapped variable and has the classical value $ m_0 = 0 $.

\section{Mean-field analysis \label{SecSummary}}

Let us for the moment neglect all quantum fluctuations and briefly repeat the
mean--field analysis which has been presented in Ref.\cite{KL-SU2-short}. The KL contains
two fermionic sectors which can have different gaps depending on the band filling
and the spin configuration. The gaps can be found from Eqs.(\ref{half_params}-\ref{hel_params}):
\bea
   \label{Gap12}
   {\cal L}_{\rm bs}^{\rm (1/2)}:     & \ &    \Delta^{(1/2)}_{1,2} = \tilde{J} ; \\
   \label{Gap14}
   {\cal L}_{\rm bs}^{\rm (1/4)}:     & \ &    \Delta^{(1/4)}_{1,2} = \tilde{J} (\cos(\alpha) \pm \sin(\alpha)) / \sqrt{2} ; \\
   \label{GapHel}
   {\cal L}_{\rm bs}^{\rm (gen)}: & \ &    \Delta^{\rm (gen)}_{1} = \tilde{J}, \ \Delta^{\rm (gen)}_{2} = 0 .
\eea
{\it At special commensurate fillings}, the minimum of the ground state energy is provided by
those spin configurations which open the gaps at the Dirac point of both fermionic sectors,
i.e., by the commensurate configurations Eq.(\ref{half_params}) and Eq.(\ref{quart_params})
[with $ \alpha = 0, \pi/2 $] for the 1/2- and 1/4-filling, respectively. Note that $ \alpha $ is gapped
at quarter-filling. The conduction band of these commensurate KLs is empty (see Fig.\ref{GS-Fig1}b)
and, hence, they  are {\it insulators}, as expected.

The commensurate spin configurations minimize the ground state energy also in {\it a vicinity of half-
and quarter fillings}. This means that the wave vector $ q $ of the spin modes remains commensurate,
Eqs.(\ref{half_params},\ref{quart_params}), and is slightly shifted from $ 2 k_F $: $ 2 k_F - q \equiv Q \ll 1/\xi$.
This case can be described in terms of backscattered Dirac fermions which, unlike the standard approach
describing 1D systems, have non-zero chemical potential:
%%%
% This yields smooth $ Q $-oscillation of the Lagrangians, e.g.:
%%%
\be
     \label{shifted_mu}
%%%
%     \mbox{near 1/2-filling}: \,
%%%
     \bar{{\cal L}} = {\cal L}_F[R,L] + {\cal L}_{\rm bs}^{(f)}[R,L] - (v_F Q / 2 ) (R^\dagger R + L^\dagger L) ,
%%%
%     {\cal L}_{Q}^{(1/2)} \!\!= \tilde{J} \left( e^{-i Q x}\tilde{R}^\dagger \s_x \tilde{L} + h.c \right) ;
%%%
\ee
where $ f = 1/4, 1/2 $. To be definite we analyze upward shift of the chemical potential, see Fig.\ref{GS-Fig1}c,
downward shift can be studied in much the same way. Backscattering caused by the commensurate spin
configuration opens the gap which is now slightly below the level of $ \mu $. The states with energies $ 0 < E \le
v_F Q / 2 $ are pushed above the gap, see Fig.\ref{GS-Fig1}d. These electrons have almost parabolic dispersion:
\be
  \label{DispRel}
  E^{+}(k) \bigl|_{v_F |k| < | \tilde{J} | } \simeq |\tilde{J}| + \left( v_F k \right)^2 / 2 |\tilde{J}| \, .
\ee
Since this new phase possesses a partially filled band it is a metal. Its metallic behavior originates from the
spin configuration whose classic component is governed by only one slowly rotating vector, e.g. $ \vec{e}_1 $,
and, therefore, is close to the collinear one. We will reflect this fact by referring to such phases as ``{\it collinear
metals}'' (CMs).

The CM becomes less favorable phase when $ | Q | $ increases. This is obvious
from Fig.\ref{GS-Fig1}: the potential energy of the electrons in the upper band
\[
 E_{\rm p} \simeq \xi \tilde{J} k_F^* / \pi + \xi v_F^2 (k_F^*)^3 / 6  \pi \tilde{J} \, ;
\]
becomes large when $ k_F^* = Q  / 2 $ increases . If $ | Q | $ is large enough
\be
    | Q | > Q_c \sim \tilde{J} / v_F\, .
\ee
the minimum of the ground state energy is provided by the general spin configuration, Eq.(\ref{hel_params}).
We would like to emphasize that the spin configuration cannot change gradually. The switching from the
commensurate to the generic configuration is always abrupt and, therefore, $ Q_c $ is the point of a
{\it quantum phase transition}.

If $ \tilde{J} / v_F \ll 1/\xi $, there is always a parametrically large window of the band fillings where
the new phase is realized instead of the CM, see the central panel of Fig.\ref{GS-NumPhDiagr}.
In the opposite case of large $ \tilde{J} $, this window shrinks to zero and the CM dominates at
all fillings excluding two special commensurate cases 1/2 and 1/4, see the lower panel of Fig.\ref{GS-NumPhDiagr}.
This allows us to surmise that the CM,
which we have predicted, corresponds to the ``polaronic liquid'' reported in Ref.\cite{mcculloch_2002}
for the case $ J_K \sim E_F $.  The detailed theory of KL with large $ J_K $ is beyond the scope of the
present paper.

The remaining case of generic filling, Eq.(\ref{hel_params}), is the most prominent for transport because rotated
fermions are gapped only in one helical sector, e.g. $\tilde{R}_{\downarrow}, \tilde{L}_{\uparrow} $, and the second
helical sector, $ \tilde{R}_{\uparrow}, \tilde{L}_{\downarrow} $, remains gapless, see Eq.(\ref{GapHel}) and Fig.\ref{GS-Fig1}e.
This means that the helical symmetry of the fermions $ \tilde{R}, \tilde{L} $ is broken on the mean-field level. On the other
hand, the rotating matrix $ g $ slowly changes in space and time and, therefore, the spin configuration
can be characterized only as a local helix.  To reflect this, we refer to the KLs with the locally broken
as {\it helical metals} (HMs). It has been shown in Ref.\cite{KL-SU2-short} that the HMs {\it inherit the symmetry protection
of those KLs where the helical symmetry is broken globally} \cite{TsvYev_2015,Schimmel_2016}, see also the
discussion in the next Section.

\section{Quantum-mechanical theory\label{QMTheor}}

Before presenting the quantum-mechanical approach, let us summarize its key steps.
Firstly, we integrate out gapped fermions, exponentiate the fermionic determinant  as
$ \rm{Tr} \log [\hat{G}_0^{-1} + \hat{\cal G} + \delta\hat{\Delta}] $, with
\be
  \hat{G}_0^{-1} = \left(
            \begin{array}{cccc}
                \p_+ &  0    &   0      &  \Delta_2  \\
                  0  & \p_+  &   \Delta_1    &    0  \\
                  0  &  \Delta_1  &   \p_-   &    0  \\
                \Delta_2  &  0    &   0      &   \p_-
            \end{array}
                    \right);
\ee
describing the fermions in the case of the classical configuration of spins. Spin and gap fluctuations
are contained in matrices $ \hat{\cal G} = {\rm diag} ( g^{-1} \p_+ g,  g^{-1} \p_- g ) $ and $ \delta\hat{\Delta} $
respectively; detailed definitions are given in \appendixname~\ref{DetExpand}. To derive the effective
theory, we add to the exponentiated determinant the Jacobian of the SU(2) rotation and expand obtained
Lagrangian in gradients of the matrix $ g $
%%%
% (more precisely, in $ \phi_\pm = g^{-1} \p_\pm g $)
%%%
and in small fluctuations of $ | m | $ around its classical value $ m_0 = 0 $. The commensurate spin configuration,
which corresponds to 1/4-filling, requires also the expansion in fluctuations of $ \alpha $.

Next, we reinstate the Wess-Zumino term for the spin field \cite{ATsBook}, which
is required by the quantum theory:
\be
   \label{WZ}
   {\cal L}_{\rm WZ} = i s \int_0^1 {\rm d}u \, (\vec{N}, [\p_u \vec{N} \times \p_\tau \vec{N}]) \, ;
\ee
where $ u $ is an auxiliary variable, $ \vec{N}(u=1) = \vec{S}/s $ and $ \vec{N}(u=0) = (1,0,0) $.
We insert the decomposition Eq.(\ref{SpinDecomp}) into Eq.(\ref{WZ}) and select non-oscillating
parts of $ {\cal L}_{\rm WZ} $. The commensurate spin configurations generate also the topological
term (see Ref.\cite{Tsvelik_1994}, Sect.16 of the book \cite{ATsBook}, and references therein).

Finally, we integrate out fluctuations of $ | m | $ (and of $ \alpha $, if needed) in the Gaussian
approximation.

These steps result in the nonlinear $ \sigma $-model (nLSM) in 1+1 dimensions, which describes
the spin degrees of freedom. We will argue that
%%%
% excitations of the $ \sigma $-model are gapped in the IR limit in all considered cases. This
% confirms stability of
%%%
the theory which we suggest is stable.
%%%
% If the phase is metallic,
% these modes are coupled to remaining itinerant fermions, see Eq.(\ref{RotatedFerm}), and
% mediate the effective interaction of the gapless fermions.
%%%

The nLSM depends on the band filling. Its derivation is rather lengthy but standard.
Therefore, we will present in the main text only answers and explain the algebra in
Appendices \ref{DetExpand}--\ref{Grad-top}.

\subsection{Insulating KL at special commensurate fillings \label{SecInsulator}}

The Lagrangian of $ \sigma $-model at the special commensurate fillings takes the following form:
%%%
% \bea
%   {\cal L}^{(1/2)} & = & \frac{\rho_0}{4}
%                   \left[
%                      \left( 1 + \frac{s^2 }{( 2 \rho_0 \xi \tilde{J})^2 L} \right) (\p_\tau \vec{e}_1)^2 +  (v_F \p_x \vec{e}_1)^2
%                   \right] + \cr
%                           & + & {\cal L}^{(1/2)}_{\rm top} \, .
% \eea
%%%
\be
  {\cal L}^{(f)} = \frac{1}{2 g_{f}}
                  \left[
                     \frac{(\p_\tau \vec{e}_1)^2}{c_{f}}  +  c_{f} (\p_x \vec{e}_1)^2
                  \right] \, ;
%%%
% + {\cal L}^{(f)}_{\rm top} \, ;
%%%
%      \cr
%      c_{1/2} & = & \frac{\rho_0 v_F^2 g_{1/2}}{2}, \,
%      g_{1/2}    =   4 \pi / \sqrt{1 + \frac{s^2 }{( 2 \rho_0 \xi \tilde{J})^2 L} } .
%     \nonumber
%%%
\ee
where ``$ f $'' denotes the band filling, either 1/2 or 1/4, $ c_{f} = v_F g_{f} / 4 \pi $ is
the normalized velocity of the spin excitations and $ g_{f} $ is the dimensionless
coupling constant:
\bea
   g_{1/2} & = & 4 \pi / \sqrt{1 + s^2 / ( \vartheta_0 \xi \tilde{J})^2 \log[D/|\tilde{J}|] } \, ; \label{gn}\cr
   g_{1/4} & = & 4 \pi / \sqrt{1 + 8 s^2 / ( \vartheta_0 \xi \tilde{J})^2 \log[\sqrt{2}D/|\tilde{J}|] } \, .
\nonumber
\eea
Clearly, the $ \sigma $-model contains only one vector, $ \vec{e}_1 $ in our choice of
variables, which governs the fermionic gap. The second vector, $ \vec{e}_2 $, is redundant in the
case of the special commensurate fillings.
%%%
% and $ \vec{m} $ does not appear in the low-energy theory because it is gapped variable.
%%%

The action is given by the sum of the gradient- and topological terms:
\be
   S^{(f)} = \int {\rm d} \tau {\rm d} x {\cal L}^{(f)} + S_{\rm top} , \ S_{\rm top} = (2 s - 1 ) i \pi k \, .
\ee
%%%
% $ S_{\rm top} $ is derived in \appendixname~\ref{WZtop} and \appendixname~\ref{Grad-top}.
%%%
The integer $ k $  marks topologically different sectors of the theory.

Hence, the spin excitations at the special commensurate cases are described by the O(3)-symmetric
nLSM in (1+1) dimensions with the topological term. Its spectrum depends on the value
of $ 2 s - 1 $: half-integer $ s $ are all equivalent to zero topological term and integer ones to $ S_{\rm top} =
i \pi$. The O(3) nonlinear sigma model is exactly solvable in both cases \cite{Wiegmann_1985,Fateev_1991,ATsBook}.
It possesses a characteristic energy scale
\be
   {\cal E}_{f} \sim | \tilde{J} | g_f^{-1}\exp(- 2\pi/g_f), \ f = 1/2, 1/4 \, .
\ee
For half-integer $s$, this scale is the spectral gap of the coherent triplet excitations whose dispersion
has a relativistic form:
\bea
   \mbox{half-integer $s$}: \quad \epsilon_f(p) = \sqrt{ {\cal E}_{f}^2 + (c_f p)^2}.
\eea
For integer $s$, $ {\cal E}_{f} $ marks a crossover from the weak coupling regime to a critical state [in the field-theoretical
language, it is described by the SU$_1$(2) Wess-Zumino-Novikov-Witten theory with central charge $C=1$]. Below $ {\cal E}_{f} $,
such KL behaves as spin-1/2 Heisenberg antiferromagnet with incoherent spin response and gapless excitation spectrum
consisting of spin-1/2 spinons.

We note that, for small $ \tilde{J} $, the energy scale $ {\cal E}_{f} $ is exponential in $1/|\tilde{J}|$ and, for
$ \tilde{J} >0 $, it may be confused with the Kondo temperature. However, as we have mentioned above, $ {\cal E}_{f} $
does not depend on the sign of $ \tilde{J} $
which testifies to the fact that the underlying physics is related to the RKKY exchange and not to the Kondo screening.

\subsection{KL in a vicinity of special commensurate fillings \label{SecHLL}}

We have explained already that the spin configuration remains unchanged if the
band filling is slightly detuned from one of the special commensurate cases. The
novel features of such KLs are caused by the presence of the conduction electrons.
They are coupled to spin modes which can mediate indirect
electron interaction. In the rotated fermionic basis, the electron-spin coupling
is described by
\be
  \label{ElSp_Int}
  {\cal L}_{\rm e-s} = \tilde{R}^\dagger g^{-1} \p_{+} g \tilde{R}
                                  + \tilde{L}^\dagger g^{-1} \p_{-} g \tilde{L} \,
\ee
(see Eq.(\ref{RotatedFerm}) and the first paragraph in \appendixname~\ref{DetExpand}). The energy of the
effective electron interaction can be estimated (at least for the case of the gapped spin excitations) after
selecting in Eq.(\ref{ElSp_Int}) the contribution of the only relevant vector $ \vec{e}_1 $
\[
  g \to i (\vec{\s},\vec{e}_1), \quad
  g^{-1} \p_\alpha g \to i \bigr( \vec{\s}, [ \vec{e}_1 \times \p_\alpha \vec{e}_1 ] \bigl) \, ;
\]
(see \appendixname~\ref{UslRel}),
neglecting time derivatives, and integrating over $ \p_x \vec{e}_1 $. This yields:
\be
  \label{ElEl_Int}
  {\cal L}^{\rm (eff)}_{\rm e-e} \sim
      - \bigl[ ( \tilde{R}^\dagger \vec{\s} \tilde{R} )^2 +  ( \tilde{L}^\dagger \vec{\s} \tilde{L} )^2 \bigr] / \vartheta_0 \, .
\ee
Thus, for the dressed fermions, we obtain a strongly repulsive Fermi gas. In the energy range close to the new
Fermi surface,  $ | E - E^+(k_F^*) | \ll E^+(k_F^*) $, the effective interaction
%%%
% \tilde{J} - v_F k_F $,
%%%
converts the conduction electrons to the repulsive and spinful TLL whose excitations are charge and spin
density waves.  This TLL is characterized by a new Fermi momentum $ k^*_F = Q / 2 = k_F - \pi / 2 \xi $ near
1/2-filling or $ k^*_F = Q / 2 = k_F - \pi / 4 \xi $ near 1/4-filling. If
%%%
% the spin modes are localized and
%%%
the effective repulsion is strong enough, TLL becomes {\it heavy}.  In particular, the velocity of CDW becomes
much smaller than $ v_F $. Such a heavy TLL has been  observed numerically in Ref.\cite{khait_2018}.

\begin{figure}[t]

   \includegraphics[width=0.4 \textwidth]{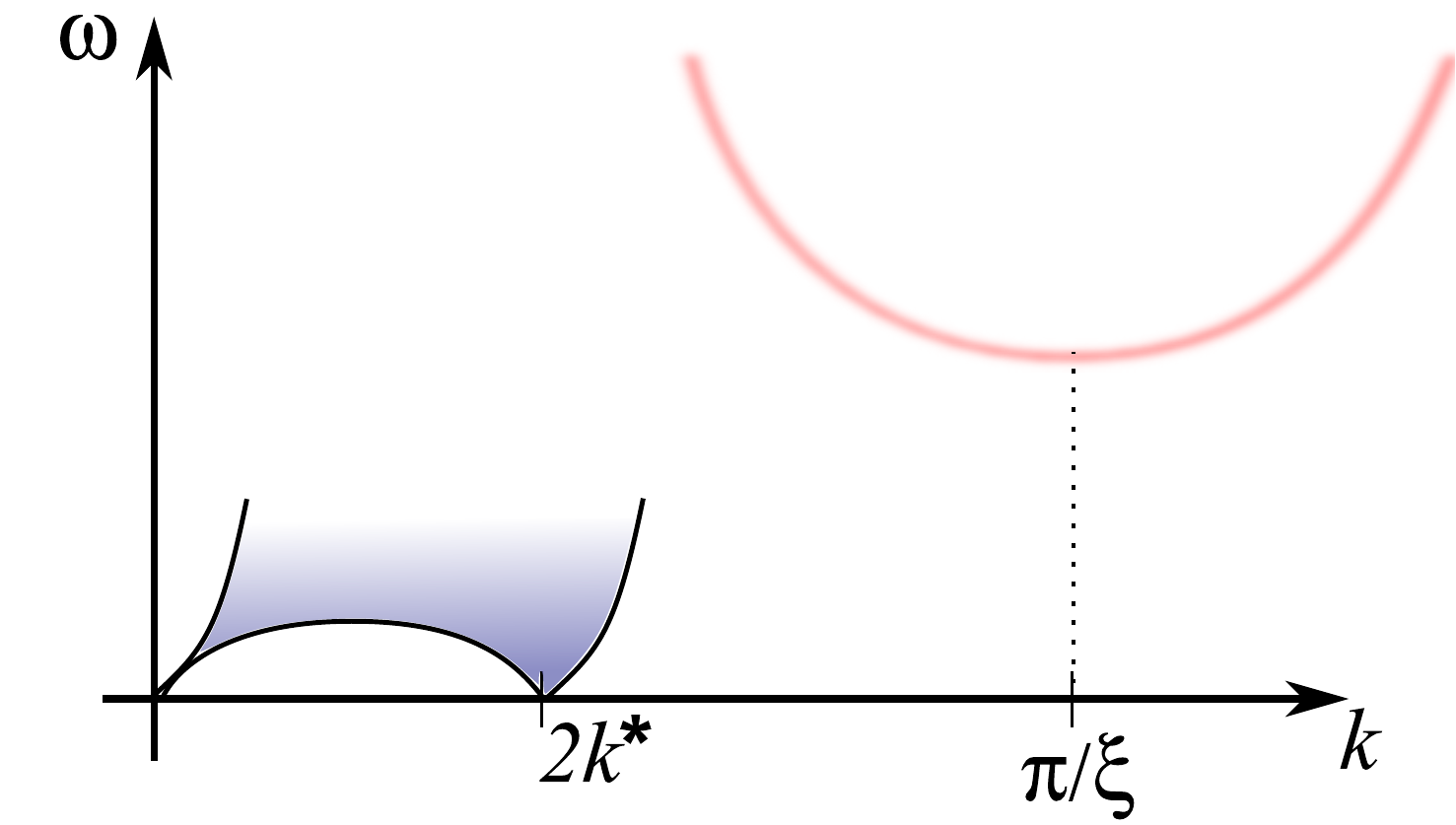}

\vspace{-0.25cm}
   \caption{
        \label{Spin-Res} (color on-line)
        The spin response of the KL close to the special commensurate fillings. The main response
        is at frequencies higher than the spin gap and coming from the vicinity of the commensurate wave vector
        (e.g., $\pi / \xi$ for half filling). The spin response of TLL  is shifted to the region of wave vectors between
        0 and $2k_F^*$
           }
\end{figure}

The spin response of the conduction electrons  is shifted to the region of wave vectors between 0 and $2k_F^*$
while the main response of the spin sector is at frequencies higher than the spin gap  coming from the vicinity
of the commensurate wave vector ($\pi / \xi$ for half filling, $\pm \pi / 2 \xi $ for quarter filling), see a cartoon in
Fig.\ref{Spin-Res}.

The appearance of the new Fermi vector $ k^*_F $ points to the existence of the large FS. The response
at $ 2 k_F $ is suppressed and is replaced by the singular response at $ 2 k^*_F $ in agreement with the
general theory \cite{yamanaka_1997,oshikawa_2000}. Note that 1D nature and the effective repulsion
make CMs very sensitive to spinless impurities: even a weak disorder easily drives it to the Anderson localized
regime with suppressed dc transport \cite{GiaSchulz}.

\subsection{KL at arbitrary fillings \label{SecHeliMet}}

The Lagrangian of the $ \sigma $-model at the generic band filling takes the following form:
\be
\label{SM-hel}
  {\cal L}^{\rm (gen)} = \frac{1}{2 g_{\rm gen}}
                  \left\{
                     \frac{ \left[ \Omega_\tau^{(z)} \right]^2 }{c_{\rm gen}}  +  c_{\rm gen} {\rm tr}( \p_x g^+ \p_x g)
                  \right\} \, ;
\ee
with
\be
   g_{\rm gen} = \pi / \sqrt{ 1 + 8 s^2 / ( \vartheta_0 \xi \tilde{J})^2 \log[D/|\tilde{J}|] }
\ee
and $ c_{\rm gen} = v_F g_{\rm gen} / \pi, \ \Omega_\tau^{(z)} \equiv
i {\rm tr}[ \s_{z} g^{-1} \p_\tau g] / 2 $. This theory is anisotropic and has only the
SU(2)-symmetry, $ g \to {\cal M} g, \, {\cal M} \in \mbox{ SU(2)} $. Moreover, the time
derivative is present only in the $\Omega^z$ term. This points to a short correlation
length of spins and may challenge our approach, which is based on  separation
of the fast and the slow degrees of freedom in the spin dynamics. However, the $ \sigma $-model
Eq.(\ref{SM-hel}) has been derived for scales larger than the coherence length of the
gapped fermions, $ v_F / \Delta^{\rm (gen)} \gg \xi $. Thus, the actual UV cut-off
of the theory is much larger than the lattice spacing and the working hypothesis on
the scale separation is not violated. The spin gap is expected to be $ \propto
\Delta^{\rm (gen)} \sim \tilde{J} $. More detailed analysis of the spin dynamics in this
generic phase will be present elsewhere.

The gapped spins mediate repulsion between
the gapless electrons. Its strength can be estimated similar to the CM case.

Let us now identify the nature of the HM in terms of its low energy excitations.
First, we note that the fermion density, current density and backscattering operators are invariant under $ g $-rotation
 (\ref{RotatedFerm}):
\be
  R^\dagger R = \tilde{R}^\dagger \tilde{R}, \ L^\dagger L = \tilde{L}^\dagger \tilde{L}, \
  R^\dagger L = \tilde{R}^\dagger \tilde{L} \, .
\ee
The low energy physics is governed by fields whose correlation functions decay as  power law.
To obtain them, we project the fields on the gapless sector, i.e., average over high-energy gapped modes.
For example, components of the charge density are:
\bea
 q \simeq 0:         & \ &
             \rho_0 = \tilde{R}^\dagger_{\uparrow} \tilde{R}_{\uparrow} + \tilde{L}^\dagger_{\downarrow} \tilde{L}_{\downarrow} \, ; \\
\label{4kF}
 q \simeq 4 k_F: & \ &
             \rho_{4 k_F} =
      e^{- 4 i k_F x} \tilde{R}^\dagger_{\uparrow} \langle \tilde{R}^\dagger_{\downarrow} \tilde{L}_{\uparrow} \rangle \tilde{L}_{\downarrow}
      + h.c.
\eea
The absence of $ \rho_{2 k_F} $ has a simple physical explanation: it would correspond to a single particle
elastic backscattering between the gapless fermion and the gapped one which is not allowed. The same
argument was used to omit contributions to $ \rho_{4 k_F} $ which are not included in Eq.(\ref{4kF}). Since
the projection of the spin density on the low energy sector vanishes, the helical metal is the $ 4 k_F $-CDW
phase. This fact has two important consequences: (i) the (local) spin helix moves the Friedel
oscillations of the charge density from $ \, 2k_F \, $ to $ \, 4k_F \, $, which is indistinguishable from
$ \, 4 (k_F - \pi/2 \xi ) \, $ due to $ k $-periodicity; and, even more importantly, (ii) it drastically reduces
backscattering caused by spinless disorder, see Ref.\cite{KL-SU2-short} for details.

\section{Discussion of further \\
numerical and experimental studies \\
of Kondo Lattices \label{SecExper}}

The insulating KLs were studied numerically in several papers, for example, in Ref.\cite{xavier_2003}
(1/4-filling) and Ref.\cite{Huang_2019} (3/4-filling).
A very important task for the subsequent research is to reliably detect different metallic phases
in 1D KL. This requires to tune system parameters, in particular, the band filling and the Kondo
coupling. Detecting the CM is a relatively simple task because it is generic at relatively
large $ J_K $ and filling away from 1/2, 1/4 and 3/4. The heavy TLL formed by the interactions
in the CM has been observed in the numerical results of Ref.\cite{khait_2018}. However, $ J_K $
was too large for finding the HM. The KL
studied in Ref.\cite{smerat_2011} exhibits an unexpected $ 2 k_F $-peak at small $ J_K $.
Yet, the peak was detected in the spin susceptibility of 6 fermions
distributed over 48 sites. So small KL cannot yield a conclusive support or disproof
of our theory. A more comprehensive study of the larger KLs is definitely needed.

We have already described conditions under which one or other metallic phase appears. Here,
we would like to recapitulate their key features which could help to distinguish these phases in numerics
and in experiments. The conductance of the CM is equal to the quantum $ G_0 = 2 e^2 / h $
while the HM must show only $ G_0 /2 $ conductance due to the lifted spin degeneracy. CM is a spinful
TLL. Its response has a peak at the shifted Fermi momentum. HM must possess the same property
\cite{yamanaka_1997,oshikawa_2000}. However, HM is $ 4 k_F $-CDW and, therefore,
$ k_F $ and $ k_F^* $ are indistinguishable on the lattice. This indicates that HM
has the response peak at $ k_F $.
%%%
% In both CM and HM is the transport is not sensitive to the spin orientation
% of incoming electrons.
%%%
Since CM responds  to scalar potentials at $2k_F^*$ and HM -- at $4k_F^* \equiv 4k_F$, the
spinless disorder potential may have a profound difference with respect to the transport in the
CM and HM phases. Namely, localization is parametrically suppressed in HM.
%%%
% Finally, transport in CM is very sensitive to the presence of the spinless impurities while transport
% in HM is (at least partially) protected from the disorder effects.
%%%

The best control of the system parameters is provided by the experimental laboratory of cold
atoms where 1D KL was recently realized \cite{Riegger_2018}. Such experiments are, probably,
the best opportunity to test our theory. However, modern solid-state technology also allows one
to engineer specific 1D KL even in solid state platforms. It looks feasible to fabricate 1D KL
in clean 1D quantum wires made, e.g., in GaAs/AlGaAs by using cleaved edge overgrowth technique
\cite{CEO} or in SiGe \cite{mizokuchi_2018}. Magnetic ad-atoms can be deposited close
to the quantum wire with the help of the precise ion beam irradiation. One can tune parameters of
these artificial KLs by changing the gate voltage, type and density of the magnetic ad-atoms and
their proximity to the quantum wire. The experiments should be conducted at low temperatures,
$ T \ll \Delta, {\cal E} $, where destructive thermal fluctuations are weak.

As far as more conventional condensed matters systems are concerned, we are aware of only one
group of candidates: the quasi-one-dimensional organic compounds  Per$_2$M(mnt)$_2$ (M  =Pt, Pd).
They are considered as realizations of weakly coupled quarter-filled S=1/2 Kondo chains
\cite{henriques_1984,henriques_1986,bourbonnais_1991,matos_1996,green_2011,pouget_2017},
although the role of the interchain coupling there is not clear. According to
Refs. \cite{henriques_1984,henriques_1986,bourbonnais_1991,matos_1996,green_2011},
(Per)$_2$[Pt(mnt)$_2$] possesses a unique combination of Spin-Peierls and CDW order parameters,
which agrees with our theory. The band in the perylene chain is quarter-filled and the
band in the Pt(mnt)$_2$ chain is half-filled \cite{henriques_1986,alcacer_1980}. The perylene
chain is metallic, and at low temperature undergoes a Peierls (CDW) transition to an insulating
state where the perylene molecules tetramerize with wave vector $q_{\rm Per} = \pi /2 \xi $. The Pt(mnt)$_2$
chain is an insulator that undergoes a spin-Peierls transition where the Pt-dithiolate molecules
dimerize with wave vector $q_{\rm Pt} = \pi/ \xi $; here the spin-1/2 Pt moments form a spin-singlet.
Remarkably, even though $q_{\rm Pt} = 2q_{\rm Per}$, diffuse x-ray scattering, specific heat, and electrical
transport measurements indicate that both the CDW and SP transitions occur at the same, or very similar
temperature \cite{gama_1993,bonfait_1993}. This observation suggests that the two chains are strongly coupled
in spite of the mismatch in q-vectors.
%%%
% It would be interesting to study doped materials to see transitions to first metallic and then $4k_F$ CDW phases.
%%%

\section{Conclusions \label{SecConcl}}

We have developed a unified theory for 1D Kondo lattice with a dense array of spins in the
regime of a small and rotationally invariant Kondo coupling, $ |J_K| \ll E_F $. The physics
of such KLs is controlled by the RKKY indirect spin interaction. This is clearly demonstrated
by the fact that their low energy properties are insensitive to the sign of the Kondo exchange.
Nevertheless, the phase diagram is quite rich. We have identified three different phases. They include
(i) the insulating phase which appears at special commensurate band filling, either 1/2, or 1/4,
3/4; (ii) spinful interacting metals which exist in the vicinity of that commensurate fillings;
and (iii) $4k_F$ Charge Density Wave phase at generic band fillings, see Fig.(\ref{GS-NumPhDiagr}).
Electron-spin interactions can convert the 2nd phase to the heavy Tomonaga-Luttinger liquids.

Spin configurations, which govern the 1st and the 2nd phases, are collinear. That's why the
second phase can be called ``collinear metal''. The spin fluctuations around these classical
arrangements are described by the well-known O(3)-symmetric nLSMs with topological terms. The
commensurate insulators and the heavy TLL appearing in the collinear metal were known before
and were described analytically or numerically, cf. Refs.\cite{Tsvelik_1994} and \cite{xavier_2003,khait_2018}.
The commensurate insulators at 1/4-filling were explored even in experiments which we have discussed
in Sect.\ref{SecExper}.

Our most intriguing finding is, probably, the $4k_F$-CDW phase. The underlying
classical spin configuration is a slowly rotating helix. That's why we have referred to this
phase as ``helical metal''. The spin fluctuations around the helix are described by the nLSM
whose solution is unknown. Nevertheless, we have argued that the spin helix is stable. Suppression
of the $ 2 k_F $ response is the direct consequence of the local spin helicity. It
parametrically suppresses effects of a spinless disorder and localization. Thus, we come across
the emergent (partial) protection of transport caused by the interactions. To the best of
our knowledge, this gives the first example of such a protection in the system where the
SU(2) (spin rotation) symmetry is present in the Hamiltonian and cannot be broken spontaneously.
Our theoretical prediction, that the backscattering is suppressed in the HMs, has
also potential  applications in nanoelectronics and spintronics.
%%%
% which are briefly discussed in Ref.\cite{KL-SU2-short}.
%%%

We believe that detecting the HM, at first, in numerical simulations and, much more importantly,
in real experiments, seems to be the task of a high importance. Our results suggest how to tune
the physical parameters, in particular the band filling and the Kondo coupling, such that the
HM could be realized.

Thus, we have not only collected many pieces of knowledge into a unified physical picture but also
gained a significant breakthrough into understanding properties of the different phases in KLs. It
would be interesting to study in the future how the direct Heisenberg interaction between the spins
could modify out theory.
\vspace{0.1 cm}

\begin{acknowledgments}
%%%
% {\bf Acknowledgments}:
%%%
We are grateful to Jelena Klinovaja for useful discussions.
A.M.T. was supported by the U.S. Department of Energy (DOE), Division of Materials Science,
under Contract No. DE-SC0012704. O.M.Ye. acknowledges support from the DFG through the
grants YE 157/2-1 and YE 157/2-2. We gratefully acknowledge hospitality of the Abdus Salam ICTP
where the part of this project was done. A. M. T. also acknowledges the hospitality of Department
of Physics of Maximilian Ludwig University where this paper was finalized.
%%%
% , and the Cluster of Excellence, Nanosystems Initiative Munich.
%%%
\end{acknowledgments}

%%%
% \newpage
%%%

\bibliography{Bibliography,KL,HeliPhys}

%%%
% \include{KL-SU2-App}
%%%

\widetext

%%%
% \newpage
%
% \begin{center}
%  {\large
%    {\bf
%     Supplemental Materials for the paper
%    }
%     ``Physics of Arbitrary Doped Kondo Lattices \ldots''
%  } \\ \vspace{0.5cm}
%     by A. M. Tsvelik and O. M. Yevtushenko
% \end{center}
%%%

\appendix

%%%
% \section{Decomposition of a normalized vector field into constant and
%              oscillating parts \label{DecompApp}}
%
% Let us consider a unit-vector field, $ \vec{s} $ with
% $ | \vec{s} | = 1 $, and single out its zero mode and $ \pm q $ components:
% \be
%  \label{FieldDecomp}
%  \vec{s} = \vec{s}_0 + \vec{s}_c \cos(q x + \theta) + \vec{s}_s \sin(q x + \theta) \, .
% \ee
% Here $ \theta $ is a constant phase shift; coefficients $ \vec{s}_{0,c,s} $ must be
% $ smooth functions on the scale of $ 1 / q $.
%%%
%% the lattice spacing.
%%%
% The normalization of $ \vec{s} $ must hold true for arbitrary $ x $. This
% {\it always} requires mutual orthogonality
% \be
%   (\vec{s}_0,\vec{s}_c) = (\vec{s}_0,\vec{s}_s) = (\vec{s}_c,\vec{s}_s) = 0 \, ;
% \ee
% and proper normalizations:
% \bea
%   \label{HelConf}
%    \mbox{generic $ q $}: & \quad & | \vec{s}_c | = | \vec{s}_s | , \
%                                        | \vec{s}_0 |^2 + | \vec{s}_c |^2 = 1 \, ; \\
%   \label{OneHalfConf}
%   \sin(q x+ \theta)=0:      & \quad & | \vec{s}_0 |^2 + | \vec{s}_c |^2 = 1 \, , \qquad
%   \mbox{ or } \quad
%   \cos(q x+ \theta)=0:    \qquad | \vec{s}_0 |^2 + | \vec{s}_s |^2 = 1 \, ; \\
%   \label{OneQuartConf}
%   e^{ i (q x+ \theta )} = \pm \frac{1 \pm i}{\sqrt{2}}:
%                                       & \quad &
%            | \vec{s}_0 |^2 + \frac{ |\vec{s}_c|^2 + |\vec{s}_s|^2}{2} = 1 \, .
% \eea
% There are no other configurations which are compatible with decomposition
% Eq.(\ref{FieldDecomp}).
%%%

\section{Useful relations \label{UslRel}}

Using the matrix identities
\be
\label{MatrId}
 \left\{
 \begin{array}{l}
  \hat{A} = A^{(j)} \sigma_j, \quad  A^{(j)} = \frac{1}{2} {\rm tr}[\sigma_j \hat{A}]; \\ \\
  {\rm tr}[ \vec{\sigma} \hat{A}^{-1} \sigma_j \hat{A}] \, {\rm tr}[ \vec{\sigma} \hat{A}^{-1} \sigma_{j'} \hat{A}] =
        4 \delta_{j,j'}
 \end{array}
 \right. \qquad
 j, j' = x, y, z.
\ee
and re-parameterizing the (real) orthogonal basis $ \vec{e}_{1,2,3} $ in terms of a matrix $ g \in \mbox{SU(2)} $:
\be
\label{BasisFromSU2-app}
   \vec{e}_{1,2,3} = \frac{1}{2} {\rm tr}[ \vec{\sigma} g \sigma_{x,y,z} g^{-1}]  \, , \quad
   \vec{e}_3 = [ \vec{e}_1 \times  \vec{e}_2 ] \, , \quad
    \sum_{a=1,2,3} ( \p_\alpha \vec{e}_a)^2 = 4 {\rm tr} [ \p_\alpha g^{-1} \p_\alpha g ] \, ;
\ee
we can re-write a scalar product $ ( \vec{\s}, e_j ) $ as follows:
\be
    \label{Vec_SU2-app}
    ( \vec{\s}, \vec{e}_{1,2} )                                 = \frac{1}{2} g \sigma_{x,y} g^{-1}
       \quad \Rightarrow \quad
    ( \vec{\s}, [ \vec{e}_{1} \pm  i \vec{e}_{1} ] )  = g \sigma_{\pm} g^{-1} \, ; \quad
     \sigma_{\pm} \equiv ( \s_x \pm i \s_y ) / 2.
\ee
One can also do an inverse step and express the SU(2) matrix via a unit vector
\be
  g = i (\vec{\s},\vec{n}), \ g^{-1} = - i (\vec{\s},\vec{n}) \,; \ | \vec{n} | = 1 \, \quad \Rightarrow \quad
  g^{-1} \p_\alpha g = i \bigr( \vec{\s}, [ \vec{n} \times \p_\alpha \vec{n} ] \bigl) \, .
\ee

Another useful quantity, which will be used below, is
\be
   \Omega^{(b)}_\alpha = \frac{i}{2} {\rm tr}[ \s_{b} g^{-1} \p_\alpha g]  , \ b = x, y, z
   \ \Rightarrow \
   g^{-1} \p_\alpha g = - i \sum_b \s_b \Omega^{(b)}_\alpha \, ;
\ee
where $ \p_\alpha $ denotes some derivative. Note, that $ \Omega^{(b)}_\alpha $ defined in such a
way are real. They can be straightforwardly related to the vectors $ \vec{e}_{1,2} $:
\be
\label{OmegaVec}
   \sum_{b \ne a} \left( \Omega^{(b)}_\alpha \right)^2 = \frac{1}{4} ( \p_\alpha \vec{e}_a)^2 \, .
\ee

\section{Jacobian of the SU(2) rotation \label{SU2-Jacobian}}

Let us derive the Jacobian of the rotation
\be
  \tilde{R} = g_+^{-1} R, \ \tilde{L} = g_-^{-1} L ;
\ee
where $ g_\pm \in \rm{SU(2)} $. Formally, it is given by the following
equality:
\be
   \int {\cal D}\{ R, L \}
   e^{
      - S_+[R,R^\dagger] - S_-[L,L^\dagger]
     } = {\cal J}[g_\pm ]
   \int {\cal D}\{ \tilde{R}, \tilde{L} \}
   e^{
      - S_+[\tilde{R},\tilde{R}^\dagger; g_+]
      - S_-[\tilde{L},\tilde{L}^\dagger; g_-]
     }
\ee
($ S_\pm $ are actions of the free chiral Dirac fermions).
The Jacobian results from the chiral anomaly and is given by the ratio of determinants
\be
   {\cal J}^{-1} = \prod_{\mu = \pm} \frac{\det(\p_\mu + g_\mu^{-1} \p_\mu g_\mu)}{\det(\p_\mu)} \, .
\ee
Here, we have introduced chiral derivatives $ \p_\pm \equiv \p_\tau \mp i v_F \p_x $.
Exponentiating determinants and Taylor-expanding the logarithm, we find
\be
\label{J-1}
   \frac{\det(\p_\mu + \phi_\mu)}{\det(\p_\mu)} =
    \exp \Bigl(
       {\rm Tr} \log [ 1 + \p_\mu^{-1} \phi_\mu ]
            \Bigr) =
    \exp \left(
       - \frac{1}{2}
       {\rm Tr} \bigl[
                \p_\mu^{-1} \phi_\mu \p_\mu^{-1} \phi_\mu
                     \bigr]
            \right) ; \quad
       \phi_\mu \equiv g_\mu^{-1} \p_\mu g_\mu \, .
\ee
The linear term is absent because we are performing the expansion around equilibrium
and all higher terms are canceled out because of the so-called loop cancellation
\cite{Giamarchi,Yurkevich-2004}.
%%%
% The latter can be checked by comparing out approach with results obtained form the
% non-Abelian bosonization \cite{james-rev_2018}.
%%%
We note that the inverse chiral derivatives are the Green's functions of the Dirac fermions
\[
  \p_{\pm}^{-1} = G_{R/L} = \frac{1}{ i \omega \mp v_F k}
\]
and rewrite Eq.(\ref{J-1}) as follows
\be
\label{J-2}
      \frac{\det(\p_\mu + \phi_\mu)}{\det(\p_\mu)} =
    \exp \left(
       - \frac{1}{2}
                {\rm Tr} \bigl[ \phi_\mu \Pi_{\mu \mu} \phi_\mu ]
             \right) =
    \exp \left(
       - \frac{1}{2}
                {\rm Tr} \bigl[ \Pi_{\mu \mu} \phi^2_\mu ]
             \right)\, .
\ee
Here $ \Pi_{\mu \mu} $ are chiral response functions:
\be
   \Pi_{\mu \mu} = \int \frac{{\rm d q} {\rm d \omega}}{ (2 \pi)^ 2}
             G_\mu( \omega, k ) G_\mu( \omega + \Omega, k + Q )  =
   \frac{\mu}{2 \pi} Q G_\mu( \Omega, Q ) \to
   \frac{i \mu}{2 \pi} \p_x \p_\mu^{-1} \, .
\ee
Note that the frequency integral must be calculated prior to the integral over momentum.
Using this equation and identities
\bea
  g_{\mu}^{-1} \p_\mu g_\mu & = & - \p_\mu g_{\mu}^{-1} g_\mu
  \ \Rightarrow \
  \left( g_\mu^{-1} \p_\mu g_\mu \right)^2 = - \p_\mu g_\mu^{-1} \p_\mu g_\mu =
  \frac{1}{2} \left( g_\mu^{-1} \p^2_\mu g_\mu + \p^2_\mu g_\mu^{-1}  g_\mu \right) \, , \\
  {\rm Tr} \left( g_\mu^{-1} \p^2_\mu g_\mu \right) & = & {\rm Tr} \left( \p^2_\mu g_\mu^{-1}  g_\mu \right) \, ;
\eea
we reduce Eq.(\ref{J-2}) to
\be
  \frac{\det(\p_\mu + \phi_\mu)}{\det(\p_\mu)} =
    \exp \left(
         \frac{i \mu }{4 \pi}
                {\rm Tr} \bigl[ \p_x g_\mu^{-1} \p_\mu g_\mu \bigr]
             \right)\, .
\ee
Thus, the Jacobian reads as
\be
      {\cal J}^{-1} =
        \exp \left(
             \frac{i}{4 \pi}
             {\rm Tr} \Bigl[ \p_x g_+^{-1} \p_+ g_+ - \p_x g_-^{-1} \p_- g_- \Bigr]
                \right)\, .
\ee
If $ g_- = g_+ = g $, the Jacobian simplifies to
\be
\label{Jacobian-Answ-Sym}
g_- = g_+ = g \ \Rightarrow \
      {\cal J}^{-1} =
        \exp \left(
             \frac{v_F}{2 \pi}
             {\rm Tr} \Bigl[ \p_x g^{-1} \p_x g  \Bigr]
                \right) =
        \exp \left(
             - \frac{1}{2} v_F^2 \vartheta_0
             {\rm Tr} \Bigl[ g^{-1} \p^2_x g  \Bigr]
                \right) \, .
\ee

\section{Integrating out gapped fermions \label{DetExpand}}

The inverse Green function of the rotated fermions $ \{ \tilde{R}, \tilde{L} \} $ can be written as $ \hat{G}^{-1}
= \hat{G}_0^{-1} + \hat{\cal G} + \delta\hat{\Delta} $ with
\be
  \hat{G}_0^{-1} = \left(
            \begin{array}{cccc}
                \p_+ &  0    &   0      &  \Delta_2  \\
                  0  & \p_+  &   \Delta_1    &    0  \\
                  0  &  \Delta_1  &   \p_-   &    0  \\
                \Delta_2  &  0    &   0      &   \p_-
            \end{array}
                    \right); \quad
  \hat{\cal G} =  \left(
            \begin{array}{cc}
                 \phi_+  &   0   \\
                    0       &   \phi_-
            \end{array}
                 \right); \quad
  \delta\hat{\Delta} = \left(
            \begin{array}{cccc}
                   0      &         0           &       0            &  \delta \Delta_2  \\
                   0      &         0           &  \delta \Delta_1  &       0            \\
                   0      &   \delta \Delta_1  &        0           &       0            \\
         \delta \Delta_2  &        0           &       0            &       0
            \end{array}
                 \right)  .
%%%
%       \quad
%       \Omega^a_{\pm} \equiv \frac{1}{2}\mbox{tr}[\s^a g^{-1}\p_\pm g].
%%%
\ee
$ \hat{G}_0 $ describes the fermions in the case of the classical configuration of spins;
$ \Delta_{1,2} $ and $ \delta \Delta_{1,2} $ are classical gap values and gap fluctuations, respectively;
$ \hat{\cal G} $ describe spin fluctuations. In the main text, we have introduced the symmetric unitary
rotation of all four fermions, Eq.(\ref{RotatedFerm}), by the single matrix $ g $.
%%%
% : $ R^\dagger g \hat{\s}_x g^{-1} L \to  \tilde{R}^\dagger \hat{\s}_x \tilde{L} $.
%%%
For completeness, we keep here different matrices $ \phi_\pm = g^{-1}_\pm \p_\pm g_\pm $ and will
reinstate the equality $ g_+ = g_- $ at a later stage of calculations.

The Green's function of each gapped fermionic sector at $ \hat{\cal G} = \delta\hat{\Delta} = 0 $ is
\be
  \hat{G}(\Delta_{1,2}) =
    \left(
      \begin{array}{cc}
         G_F^{(+)}  &      G_B    \\
           G_B      &   G_F^{(-)}
      \end{array}
    \right);
  \quad
    G_F^{(\pm)} = \frac{i \omega \pm v_F q}{\omega^2 + (v_F q)^2 + \Delta_{1,2}^2} , \
    G_B = \frac{\Delta_{1,2}}{\omega^2 + (v_F q)^2 + \Delta_{1,2}^2} .
\ee

We have to integrate out gaped fermions, exponentiate the fermionic determinant as $ \rm{Tr} \log [\hat{G}_0^{-1}
+ \hat{\cal G} + \delta\hat{\Delta}]$ and expand it in all fluctuations. After Taylor-expanding logarithm, we trace out
high-energy degrees of freedom and obtain low-energy Lagrangians.

The expansion in the gap fluctuations yields
\be
\label{Lg}
  {\cal L}_g \simeq - \sum_{l=1,2} {\rm Tr} \Bigl( G_B[\Delta_l] \Bigr) \delta \Delta_l =
                    - \frac{\vartheta_0}{2} \sum_{l=1,2} \Bigl[ \Delta_l \log(D / |\Delta_l|) \, \delta \Delta_l \Bigr] \, ;
\ee
where
\bea
   \mbox{1/2-filling:} \
    \Delta_{1,2} & = & \tilde{J} \, , \ \delta \Delta_{1,2} = - \frac{1}{2} \tilde{J} m^2 \, ; \cr
   \mbox{1/4-filling:} \
    \Delta_{1,2} & = & \frac{1}{\sqrt{2}} \tilde{J} \, , \
          \delta \Delta_{1,2} = - \frac{1}{2\sqrt{2}} \tilde{J} ( m^2 + \alpha^2 \pm 2 \alpha ) \, ; \cr
   \mbox{generic filling:} \
    \Delta_1 & = & 0 \, , \ \Delta_{2} = \tilde{J} \, , \ \delta \Delta_{2} = - \frac{1}{2} \tilde{J} m^2 \, .
   \nonumber
\eea

The expansion in the spin fluctuations is done at $ \delta \Delta_{1,2} = 0 $. $ \rm{Tr} \log [\hat{G}_0^{-1} ] $
determines the ground states energy and, therefore, the linear terms in expansion in the spin fluctuations are absent:
\be
\label{SM-1}
  {\rm Tr} \log [1 + \hat{G}_0 \hat{\cal G} ] \simeq - \frac{1}{2} {\rm Tr}
                  \left[ \hat{G}_0 \hat{\cal G} \hat{G}_0 \hat{\cal G} \right] .
\ee
Similar to the derivation of the Jacobian, Eq.(\ref{SM-1}) can be rewritten in terms of the response
functions. The difference is that they are now short-range response functions of the gapped fermions.
Since typical energy scales of the spin fluctuations are expected to be much smaller than gaps, the
response functions can be calculated at zero frequency and momentum.

The contribution to the action generated by the gradients of $ g $ (i.e., by the spin fluctuations)
reads as
\be
  \delta S_g = \frac{1}{2} {\rm Tr}
                  \left[ \hat{G}_0 \hat{\cal G} \hat{G}_0 \hat{\cal G} \right] - \log({\cal J}) \, .
\ee

\subsection{The case of the commensurate insulator}

Let us at first consider the case $ \Delta_1 = \Delta_2 $, i.e. all four fermionic modes are gapped, where we obtain:
\bea
\label{SM-2}
      \hat{G}_0 & = & G_F^{(+)}  ( \hat{\tau}_+  \hat{\tau}_- \times \hat{\s}_0 )
                               + G_F^{(-)}  ( \hat{\tau}_-  \hat{\tau}_+ \times \hat{\s}_0 )
                               + G_B ( \hat{\tau}_1 \times \hat{\s}_x )
          \quad \Rightarrow \\
    - \frac{1}{2} {\rm Tr}
                  \left[ \hat{G}_0 \hat{\cal G} \hat{G}_0 \hat{\cal G} \right] & = &
    - \frac{1}{2} {\rm Tr}
                  \left[
     G_F^{(+)} \phi_+ G_F^{(+)} \phi_+ + G_F^{(-)} \phi_- G_F^{(-)} \phi_- +
     G_B \hat{\s}_x \phi_+ G_B \hat{\s}_x \phi_- + G_B \hat{\s}_x \phi_- G_B \hat{\s}_x \phi_+
                  \right] .
\eea
Here $ \hat{\tau}_j $ are the Pauli matrices which operate in the chiral sub-space.
The response functions which we need are
\be
\label{SM-3}
  \Pi_{FF} = \int \frac{ {\rm d}q   {\rm d}\omega}{(2 \pi)^2} \left[ G_F^{(\mu)}(\omega,q) \right]^2 = - \frac{\vartheta_0}{4} \, ; \quad
  \Pi_{BB} =  \int \frac{ {\rm d}q   {\rm d}\omega}{(2 \pi)^2} \left[ G_B(\omega,q) \right]^2 = \frac{\vartheta_0}{4} \, .
\ee
Note that the frequency integral must be calculated prior to the integral over momentum. Inserting
Eq.(\ref{SM-3}) into Eq.(\ref{SM-2}) we arrive at
\bea
\label{SM-4}
    - \frac{1}{2} {\rm Tr}
                  \left[ \hat{G}_0 \hat{\cal G} \hat{G}_0 \hat{\cal G} \right] & = &
      \frac{\vartheta_0}{8} {\rm Tr}
                  \left[
                        \phi_+ \phi_+ + \phi_- \phi_- - 2 \hat{\s}_x \phi_+ \hat{\s}_x \phi_-
                  \right] = \cr
      & = &
      \frac{\vartheta_0}{8} {\rm Tr}
                  \left[
                       g_+^{-1} \p_+^2 g_+ + g_-^{-1} \p_-^2 g_-
                       - 2 \hat{\s}_x g_+^{-1} \p_+ g_+ \hat{\s}_x g_-^{-1} \p_- g_-
                  \right]  = \cr
      & = &
      \frac{\vartheta_0}{4}
        \left\{
             {\rm Tr}
                  \left[
                      g^{-1} \p_\tau^2 g - v_F^2 g^{-1} \p_x^2 g
                  \right]
              - {\rm Tr}
                  \left[
                       \left( \hat{\s}_x g^{-1} \p_\tau g \right)^2
%%%
% \hat{\s}_x g^{-1} \p_\tau g
%%%
                       + \left( v_F \hat{\s}_x g^{-1} \p_x g \right)^2
                  \right]
          \right\}.
\eea
We use
%%%
% $ g_+ = g_- = g $
%%%
Eq.(\ref{Jacobian-Answ-Sym}) for the Jacobian (the anomalous contribution) of the symmetric rotation
with $ g_+ = g_-  = g $ and find
\be
\label{SM-5}
    - \frac{1}{2} {\rm Tr}
                  \left[ \hat{G}_0 \hat{\cal G} \hat{G}_0 \hat{\cal G} \right] + \log({\cal J}) =
      \frac{\vartheta_0}{4}
          \Bigl\{
              {\rm Tr}
                  \left[
                      g^{-1} \p_\tau^2 g + v_F^2 g^{-1} \p_x^2 g
                  \right]
              - {\rm Tr}
                  \left[
                       \left( \hat{\s}_x g^{-1} \p_\tau g \right)^2
%%%
% \hat{\s}_x g^{-1} \p_\tau g
%%%
                       + \left( v_F \hat{\s}_x g^{-1} \p_x g \right)^2
                  \right]
        \Bigr\} .
\ee
It is instructive to re-write Eq.(\ref{SM-5}) in terms of $ \Omega_{\alpha}^{(b)} $:
\bea
   {\rm Tr} \left[ g^{-1} \p_\alpha^2 g \right] = {\rm Tr} \left[ \left( g^{-1} \p_\alpha g \right)^2 \right] & = &
      - {\rm Tr} \left[ \left(\sum_b \s_b \Omega^{(b)}_\alpha\right)^2 \right] =
   - 2 {\rm Tr} \left[ \left( \Omega^{(x)}_\alpha \right)^2 + \left( \Omega^{(y)}_\alpha \right)^2 + \left( \Omega^{(z)}_\alpha \right)^2 \right] \, ; \cr
      {\rm Tr} \left[ \left( \hat{\s}_x g^{-1} \p_\alpha g \right)^2 \right] & = &
      - {\rm Tr} \left[ \left( \hat{\s}_x \sum_b \s_b \Omega^{(b)}_\alpha\right)^2 \right] =
   - 2 {\rm Tr} \left[ \left( \Omega^{(x)}_\alpha \right)^2 - \left( \Omega^{(y)}_\alpha \right)^2 - \left( \Omega^{(z)}_\alpha \right)^2 \right] \, ; \cr
     \Rightarrow \quad
    - \frac{1}{2} {\rm Tr}
                  \left[ \hat{G}_0 \hat{\cal G} \hat{G}_0 \hat{\cal G} \right] + \log({\cal J}) & = & - \frac{\vartheta_0}{2}
              {\rm Tr}
                  \left[
                      \left( \Omega^{(y)}_\tau \right)^2 + \left( \Omega^{(z)}_\tau \right)^2
                   + \left( v_F \Omega^{(y)}_x \right)^2 + \left( v_F \Omega^{(z)}_x \right)^2
                  \right] \, .
\eea
Thus, the contribution to the Lagrangian reads as:
\be
\label{SM-6}
\mbox{Special commensurate cases:} \quad
    \delta {\cal L}_g    =
             \frac{\vartheta_0}{8}
                  \left[
                     (\p_\tau \vec{e}_1)^2 +  (v_F \p_x \vec{e}_1)^2
                  \right] \, .
\ee
Here,
%%%
% $ \delta {\cal L}_g $ is the contribution to the action generated by the gradients of $ g $, and
%%%
we have used the identity Eq.(\ref{OmegaVec}). The symmetry of the theory Eq.(\ref{SM-6}) is reduced from the
SU(2)$ \times $SU(2) symmetry of the initial model, Eq.(\ref{Lbs}), to the $ O(3) $-symmetry.
%%%
% $ \times ( {\rm SU(2)} / \s_y \times \s_z ) $ because matrix $ g $ can be multiplied on the left by a constant SU(2)-matrix
% and on the right only by those matrices which commute with $ \s_x $.
%%%
This reflects the properties of Eq.(\ref{half_params}) obtained after selecting non-oscillating terms in backscattering.
Interestingly, neither the part generated by the gradient expansion nor the Jacobian are Lorenz invariant but
the Lorenz invariance is restored in the final answer Eq.(\ref{SM-6}) after summing all parts.

\subsection{The case of the helical metal}

Consider now the case where only one helical sector is gapped with the second remaining gapless: $ \Delta_1 = 0,
\Delta_2 \ne 0 $. Since we integrate our only gapped fermions, Eq.(\ref{SM-2}) must be modified by excluding gapless
modes from calculations:
\bea
\label{SM-2-hel}
      \hat{G}_0 & \to & G_F^{(+)}  ( \hat{\tau}_+  \hat{\tau}_- \times \hat{\s}_+ \hat{\s}_- )
                               + G_F^{(-)}  ( \hat{\tau}_-  \hat{\tau}_+ \times \hat{\s}_- \hat{\s}_+ )
                               + G_B ( \hat{\tau}_+ \times \hat{\s}_+ + \hat{\tau}_- \times \hat{\s}_- )
          \quad \Rightarrow \\
    - \frac{1}{2} {\rm Tr}
                  \left[ \hat{G}_0 \hat{\cal G} \hat{G}_0 \hat{\cal G} \right] & = &
    - \frac{1}{2} {\rm Tr}
                  \left[
     G_F^{(+)} \hat{\s}_+ \hat{\s}_- \phi_+ G_F^{(+)} \hat{\s}_+ \hat{\s}_- \phi_+ +
     G_F^{(-)} \hat{\s}_- \hat{\s}_+ \phi_- G_F^{(-)} \hat{\s}_- \hat{\s}_+ \phi_- +
     2 G_B \hat{\s}_- \phi_+ G_B \hat{\s}_+ \phi_-
%%%
% + G_B \hat{\s}_x \phi_- G_B \hat{\s}_x \phi_+
%%%
                  \right] = \cr
\label{SM-3-hel}
    & = & \frac{\vartheta_0}{8}  {\rm Tr}
                  \left[
                     \hat{\s}_+ \hat{\s}_- \phi_+ \hat{\s}_+ \hat{\s}_- \phi_+ +
                     \hat{\s}_- \hat{\s}_+ \phi_- \hat{\s}_- \hat{\s}_+ \phi_- -
                    2 \hat{\s}_- \phi_+ \hat{\s}_+ \phi_-
                  \right] .
\eea
Note that we have neglected the coupling of $ \phi $ to products of two fermionic fields
with different helicity (i.e., products of one gapped and one gapless fermions). One can
check that
\be
\label{SM-4-hel}
    {\rm Tr}
                  \left[
                       \hat{\s}_+ \hat{\s}_- \phi_\nu \hat{\s}_+ \hat{\s}_- \phi_\nu
                  \right] =
    {\rm Tr}
                  \left[
                     \hat{\s}_- \hat{\s}_+ \phi_\nu \hat{\s}_- \hat{\s}_+ \phi_\nu
                  \right] =
    - {\rm Tr}
                  \left[
                    \hat{\s}_- \phi_\nu \hat{\s}_+ \phi_\nu
                  \right] \, .
\ee
Therefore, only time derivatives remain in Eq.(\ref{SM-3-hel}):
\bea
\label{SM-5-hel}
    - \frac{1}{2} {\rm Tr}
                  \left[ \hat{G}_0 \hat{\cal G} \hat{G}_0 \hat{\cal G} \right] & = &
       \frac{\vartheta_0}{8}  {\rm Tr}
                  \left[
                     4 \hat{\s}_+ \hat{\s}_- \phi_\tau \hat{\s}_+ \hat{\s}_- \phi_\tau
                  \right] =
       \frac{\vartheta_0}{8}  {\rm Tr}
                  \left[
                     (\hat{\s}_0 + \hat{\s}_z ) \phi_\tau ( \hat{\s}_0 + \hat{\s}_z ) \phi_\tau
                  \right] = \cr
       & = & \frac{\vartheta_0}{8}  {\rm Tr}
                  \left[
                     \left( g^{-1} \p_\tau g \right)^2 + \left( \hat{\s}_z g^{-1} \p_\tau g \right)^2
                  \right]
\eea
Adding the Jacobian, Eq.(\ref{Jacobian-Answ-Sym}), we obtain:
\be
\label{SM-6-hel}
    - \frac{1}{2} {\rm Tr} \left[ \hat{G}_0 \hat{\cal G} \hat{G}_0 \hat{\cal G} \right] + \log({\cal J}) =
      \frac{\vartheta_0}{8}
          \Bigl\{
              {\rm Tr}
                  \left[
                      g^{-1} \p_\tau^2 g + ( 2 v_F )^2 g^{-1} \p_x^2 g
                  \right]
                + {\rm Tr}
                  \left[
                       \left( \hat{\s}_z g^{-1} \p_\tau g \right)^2
                  \right]
        \Bigr\} .
\ee
Similar to Eq.(\ref{SM-6}), we can rewrite
\bea
    {\rm Tr}
                  \left[
                       \left( \hat{\s}_z g^{-1} \p_\alpha g \right)^2
                  \right] & = &
 - {\rm Tr} \left[ \left( \hat{\s}_z \sum_b \s_b \Omega^{(b)}_\alpha\right)^2 \right] =
   - 2 {\rm Tr} \left[ - \left( \Omega^{(x)}_\alpha \right)^2 - \left( \Omega^{(y)}_\alpha \right)^2 + \left( \Omega^{(z)}_\alpha \right)^2 \right] \, ;
  \\
  \Rightarrow
               {\rm Tr}
                  \left[
                      g^{-1} \p_\tau^2 g
                  \right]
                & + & {\rm Tr}
                  \left[
                       \left( \hat{\s}_z g^{-1} \p_\tau g \right)^2
                  \right] =
                  - 4  {\rm Tr}  \left( \Omega^{(z)}_\tau \right)^2 \, .
%%%
%  =
%                     {\rm Tr} \left\{
%                          - \frac{1}{2} \sum_{j=1,2,3} ( \p_\tau \vec{e}_j )^2 + ( \p_\tau \vec{e}_3 )^2
%                                   \right\} \, .
%%%
\eea
%%%
% where $ \vec{e}_3 \equiv [ \vec{e}_1 \times \vec{e}_2 ] $.
%%%
Hence, we arrive at the answer
\be
\label{SM-7-hel}
%%%
%       - \frac{1}{2} {\rm Tr} \left[ \hat{G}_0 \hat{\cal G} \hat{G}_0 \hat{\cal G} \right] + \log({\cal J}) =
%%%
\mbox{Generic case:} \quad
    \delta {\cal L}_g =
                      \frac{\vartheta_0}{2}
                             \left\{
                               \left( \Omega^{(z)}_\tau \right)^2 + {\rm Tr} | v_F \p_x g |^2
%%%
%                         - \frac{1}{2} \sum_{j=1,2,3} ( \p_\tau \vec{e}_j )^2  + ( \p_\tau \vec{e}_3 )^2
%%%
                                  \right\} \, .
\ee

%%%
% The theory Eq.(\ref{SM-7-hel}) is anisotropic and does not have the Lorenz invariance. We will show in
% \appendixname\ref{RG-flow} that the anisotropy is irrelevant and the Lorenz invariance is restored in the
% IR limit due to the RG flow.
%%%

\section{Smooth parts of the Wess-Zumino term \label{WZsmooth}}

Let us now project the Wess-Zumino term of the action on the low-energy sector. Firstly, we
consider a given site, $ n $, use the standard expression for the Wess-Zumino Lagrangian on
this site
\be
   \label{L_WZ}
   {\cal L}_{\rm WZ} = i s \int_0^1 {\rm d} u \, ( \vec{N}, [ \p_u \vec{N} \times \p_\tau \vec{N} ] ) \, ; \
   \vec{N}(u=0) = (1,0,0), \, \vec{N}(u=1) = \vec{S}_n / s;
\ee
substitute the decomposition Eq.(\ref{SpinDecomp}) for $ \vec{N} $, and neglect all $ q $-oscillations.
Following the procedure, which is explained in detail in Sect.16 of the book \cite{ATsBook}, we arrive
at the following answer for the smooth contribution of $  {\cal L}_{\rm WZ} $:
\be
\label{WZsm}
   {\cal L}_{\rm WZ}^{\rm (sl)} = - \frac{i s b^2 (1 - m^2 )}{2} \Bigl(
                        \cos^2(\alpha) ( \vec{m}, [ \vec{e}_1 \times \p_\tau \vec{e}_1 ] )+
                        \sin^2(\alpha) ( \vec{m}, [ \vec{e}_2 \times \p_\tau \vec{e}_2 ] )
                                                                                \Bigr) \, .
\ee
If $ | \vec{m} | \ll 1 $, we must keep in Eq.(\ref{WZsm}) only terms of order $ O(m) $ and substitute
the classical value for $ \alpha $, e.g. $ \alpha = 0 $, at the 1/4-filling. Hence, Eq.(\ref{WZsm}) reduces to
\bea
%%%
% \label{WZ12}
%    {\cal L}_{\rm WZ}^{(1/2)} & = & - \frac{i s}{2}
%                        ( \vec{m}, [ \vec{e}_1 \times \p_\tau \vec{e}_1 ] )
%%%
%                         \sin^2(\alpha) ( \vec{m}, [ \vec{e}_2 \times \p_\tau \vec{e}_2 ] )
%%%
%                                                              \, ; \\
%%%
\label{WZcom}
    2 {\cal L}_{\rm WZ}^{(1/2)} & = & {\cal L}_{\rm WZ}^{(1/4)} = - i s
                        ( \vec{m}, [ \vec{e}_1 \times \p_\tau \vec{e}_1 ] )
                                                              \, ; \\
\label{WZhel}
   {\cal L}_{\rm WZ}^{\rm (gen)} & = & - \frac{i s}{2}
                        ( \vec{m}, [ \vec{e}_1 \times \p_\tau \vec{e}_1 ] +  [ \vec{e}_2 \times \p_\tau \vec{e}_2 ] )
                                                             \, ;
\eea
in the cases of the commensurate fillings, Eq.(\ref{half_params},\ref{quart_params}), and  the generic
filling, Eq.(\ref{hel_params}), respectively.

$ | \vec{m} | $ is the gapped variable. Its deviations from the classical value $ m = 0 $ reflect
the gap fluctuations and are described by the quadratic Lagrangian
\be
   {\cal L}_g = \frac{1}{2 {\cal M}^{(f)}} | \vec{m} |^2 \, ;
\ee
see Eq.(\ref{Lg}) which defines the value of the variance $ {\cal M}^{(f)} \propto 1 / \left( \vartheta_0 \tilde{J}^2
\right) $ with the superscript ``$ f $'' marking the band filling. Now, we have to calculate the Gaussian
integral over $ \vec{m} $. This step depends on the spin configuration

\subsection{Cases of the special commensurate band filling}

Eq.(\ref{WZcom}) includes only one vector, $ \vec{e}_1 $. Therefore, the direction of the vector $ \vec{m} $
is not fixed by the effective action and  one must integrate over  $ |\vec{m}| $ and over all directions of this
vector. In the Gaussian approximation, this results in:
\be
\label{Int-m_Commens}
  {\cal N} \int_{-\infty}^{+\infty} {\rm d} m_{x,y,z} e^{- \int {\rm d}\{\tau,x\} \left[ {\cal L}_g + (\vec{m}, \vec{A} ) \right] } =
  \exp\left[ \frac{{\cal M}^{(f)}}{2}  ( \vec{A}, \vec{A} ) \right] \, ;
\ee
with $ {\cal N} $ being the normalization factor. Eq.(\ref{Int-m_Commens}) yields contributions to the Lagrangian,
which are governed by the smooth part of the Wess-Zumino term in the special commensurate cases:
\bea
\label{WZ-12}
   \delta {\cal L}_{\rm WZ}^{(1/2)} & = & \frac{s^2 }{8 \, \vartheta_0 (\xi \tilde{J})^2 \log(D/|\tilde{J}|)} (\p_\tau  \vec{e}_1)^2
%%%
%               = \frac{s^2 }{8} {\cal M}^{(1/2)} \left[ \left( \Omega_\tau^{(y)} \right)^2 + \left( \Omega_\tau^{(z)} \right)^2 \right]
%%%
          \, ; \\
%%%
%            \quad
%           1/{\cal M}^{(1/2)} = 2 \rho_0 \tilde{J}^2 \log(D/|\tilde{J}|) \, ; \\
%%%
\label{WZ-14}
   \delta {\cal L}_{\rm WZ}^{(1/4)} & = & \frac{s^2 }{\vartheta_0 (\xi \tilde{J})^2 \log(\sqrt{2}D/|\tilde{J}|) } (\p_\tau  \vec{e}_1)^2  \, ; \\
%%%
%           \quad
%           1/{\cal M}^{(1/4)} = \xi \rho_0 \tilde{J}^2 \log(\sqrt{2}D/|\tilde{J}|) \, ; \\
%%%
\eea
Here, we have used the identity $ [ \vec{e} \times \p  \vec{e}]^2 = (\p  \vec{e})^2, | \vec{e} | = 1 $.
Adding Eqs.(\ref{SM-6}) and either (\ref{WZ-12}) or (\ref{WZ-14}) yields the $ \sigma $-model describing
the spin excitation at the special commensurate filling (1/2 and 1/4 respectively) and in its vicinity.

\subsection{The case of the generic band filling}

Unlike the special commensurate cases, Eq.(\ref{WZhel}) contains both vectors, $ \vec{e}_{1,2} $. Therefore, the direction
of the vector $ \vec{m} $ is fixed: $ \vec{m} = | \vec{m} | [ \vec{e}_1 \times \vec{e}_2 ] $ and one must integrate {\it only} over
$ |\vec{m}| $. In the Gaussian approximation, this results in:
\be
\label{Int-m_Hel}
  {\cal N} \int_{-\infty}^{+\infty} {\rm d} m e^{- \int {\rm d}\{\tau,x\} \left[ {\cal L}_g + m ( \vec{e}_3, \vec{A} ) \right] } =
  \exp\left[ \frac{{\cal M}^{(f)}}{2}  ( \vec{e}_3, \vec{A} )^2 \right] \, ; \quad
  \vec{e}_3 \equiv [ \vec{e}_1 \times \vec{e}_2 ] \, .
\ee
Eq.(\ref{Int-m_Hel}) yields the contribution to the Lagrangian, which are governed by the smooth part
of the Wess-Zumino term in the case of the generic band filling:
\be
\label{WZ-gen}
   \delta {\cal L}_{\rm WZ}^{\rm (gen)} = \frac{s^2}{\vartheta_0 (\xi \tilde{J})^2 \log(D/|\tilde{J}|) }
                                  \Bigl(
                      \vec{e}_3,
                      [ \vec{e}_1 \times \p_\tau \vec{e}_1 ]
%%%
% + [ \vec{e}_2 \times \p_\tau \vec{e}_2 ] )
%%%
                                  \Bigr)^2 =
%%%
%           , \quad
%           1/{\cal M}^{\rm (gen)} = \xi \rho_0 \tilde{J}^2 \log(D/|\tilde{J}|) \, .
           \frac{4 s^2}{\vartheta_0 (\xi \tilde{J})^2 \log(D/|\tilde{J}|) } \left( \Omega_\tau^{(z)} \right)^2 \, .
%%%
\ee
Here, we have used the identity $ ( \vec{e}_3, [ \vec{e}_1 \times \p_\tau \vec{e}_1 ] ) = ( \vec{e}_3, [ \vec{e}_2 \times
\p_\tau \vec{e}_2 ] ) $. Adding Eqs.(\ref{SM-7-hel}) and (\ref{WZ-gen}) yields the $ \sigma $-model describing the spin excitation at the generic band filling.

\section{Topological part of the Wess-Zumino term \label{WZtop}}

Let us now focus on the special commensurate fillings and single out leading in $ | \vec{m} | $
parts of the Wess-Zumino Lagrangian which are sign-alternating and produce the topological
contribution to the effective action.

\subsection{Topological term for the 1/2-filling}

Following Ref.\cite{Tsvelik_1994}, we insert into Eq.(\ref{L_WZ}) only the term $ (-1)^n \vec{e}_1 $ from
Eq.(\ref{SpinDecomp},\ref{half_params}).  This yields:
\be
   \sum_n {\cal L}^{(1/2)}_{\rm top}(n) =  i s \sum_n (-1)^n\int_0^1 {\rm d} u \, ( \vec{e}_1(n), [ \p_u \vec{e}_1(n) \times \p_\tau \vec{e}_1(n) ] ) \, .
%%%
%                                                    ( 1 + \vec{m}^2 )^{3/2} \, .
%%%
\ee
After converting the sum over the lattice sites to the space integral, we obtain the topological contribution to the action:
\be
    S^{(1/2)}_{\rm top} = i \frac{s}{2} \int_0^\beta {\rm d} \tau \int_0^L {\rm d} x \, ( \vec{e}_1, [\p_x \vec{e}_1 \times \p_\tau \vec{e}_1 ] ) =
                                       2 s i \pi k \, ;
\ee
see details in Sect.16 of the book \cite{ATsBook}. The integer $ \, k \, $ marks topologically different sectors of the theory.

\subsection{Topological term for the 1/4-filling}

Similar to the previous case, we insert into Eq.(\ref{L_WZ}) only the term $ \sqrt{2} \Big( \vec{e}_1 \cos(\alpha)
\cos(\pi n / 2 + \pi/4) + \vec{e}_2 \sin(\alpha) \sin(\pi n / 2 + \pi / 4) \Bigr) $ from Eq.(\ref{SpinDecomp},\ref{quart_params}).
Using the classical value of $ \alpha $, we obtain
\bea
   \sum_n {\cal L}^{(1/4)}_{\rm top}(n) & = &  i s \sum_{n=1}^{N} (\sqrt{2} \cos(\pi n / 2 + \pi/4))
                             \int_0^1 {\rm d} u \, ( \vec{e}_1(n), [ \p_u \vec{e}_1(n) \times \p_\tau \vec{e}_1(n) ] ) \simeq \cr
        & \simeq & 2  i s \sum_{n=1}^{N/4} \int_0^1 {\rm d} u \Bigl\{
                                  ( \vec{e}_1, [ \p_u \vec{e}_1 \times \p_\tau \vec{e}_1 ] ) |_{2(2n+1)} -
                                  ( \vec{e}_1, [ \p_u \vec{e}_1 \times \p_\tau \vec{e}_1 ] ) |_{2n}
                                                                                               \Bigr\} \, ;
\eea
and find the topological contribution to the action
\be
    S^{(1/4)}_{\rm top} = i s \int_0^\beta {\rm d} \tau \int_0^{L/2} {\rm d} x \, ( \vec{e}_1, [\p_x \vec{e}_1 \times \p_\tau \vec{e}_1 ] ) =
                                       2 s i \pi k \, .
\ee

Thus, we have found $ S^{(1/2)}_{\rm top} = S^{(1/4)}_{\rm top} = 2 s i \pi k $.

\section{Topological part of the fermionic determinant \label{Grad-top}}

The gradient expansion Eq.(\ref{SM-1}) can generate subleading terms:
\be
\label{SM-1-Subl}
  {\rm Tr} \log [1 + \hat{G}_0 \hat{\cal G} ] \simeq - \frac{1}{2} {\rm Tr}
                  \left[ \hat{G}_0 \hat{\cal G} \hat{G}_0 \hat{\cal G} \right] +\frac{1}{3} {\rm Tr}
                  \left[ \hat{G}_0 \hat{\cal G} \hat{G}_0 \hat{\cal G} \hat{G}_0 \hat{\cal G} \right] .
\ee
If the fermions were gapless, all subleading terms beyond the second order in $ \hat{\cal G} $ would disappear
because of the loop cancellation \cite{Giamarchi,Yurkevich-2004}. However, the gap makes them finite, in
particular:
\bea
\label{Grad-third}
   {\rm Tr} \left[ \hat{G}_0 \hat{\cal G} \hat{G}_0 \hat{\cal G} \hat{G}_0 \hat{\cal G} \right] & = &
      {\rm Tr} \left[ ( G_F^{(+)} \phi_+)^3 +  ( G_F^{(-)} \phi_-)^3 +
                    \right. \\
&  \qquad & + \left.
                     3 G_B \hat{\s}_x \phi_+ G_F^{(+)} \phi_+ G_B \hat{\s}_x \phi_- +
                     3 G_B \hat{\s}_x \phi_- G_F^{(-)} \phi_- G_B \hat{\s}_x \phi_+
%%%
%                              G_F^{(+)} \phi_+ G_B \hat{\s}_x \phi_- G_B \hat{\s}_x \phi_+ +
%                              G_B \hat{\s}_x \phi_+ G_F^{(+)} \phi_+ G_B \hat{\s}_x \phi_- +
%                              G_B \hat{\s}_x \phi_- G_B \hat{\s}_x \phi_+ G_F^{(+)} \phi_+
%%%
                        \right] \ne 0 \, .
\nonumber
\eea
Eq.(\ref{Grad-third}) describes a nonlinear response of the system and it may contain topological
terms \cite{TITS-book}. Extracting the topological part of the nonlinear response function is a lengthy
and non-trivial task which is beyond the scope of the present paper. Instead of the direct algebra,
we will rely on the claim of Ref.\cite{Tsvelik_1994}: the topological part of the fermionic determinant
reduces the spin value in $ S^{(f)}_{\rm top} $ by 1/2. This reflect the strong coupling of the itinerant
electrons with the localized magnetic moment in the commensurate cases. Thus, we use the expression
\be
   S_{\rm top} = (2 s - 1) i \pi k \, ,
\ee
for the analysis of the special commensurate fillings.

\end{document}